\documentclass[review]{elsarticle}

\usepackage{lineno,hyperref}
\modulolinenumbers[5]
\usepackage{graphicx}                    
\usepackage{color}                       
\usepackage{breakurl}                    
\usepackage{amssymb}
\usepackage{amsbsy}
\usepackage{latexsym}
\usepackage{marvosym}
\usepackage{enumerate}
\usepackage{textcomp}
\usepackage{float}
\usepackage{times}
\usepackage[varg]{txfonts}

\renewcommand{\vec}[1]{{\boldsymbol #1}}

\journal{New Astronomy}

\usepackage{numcompress}\biboptions{authoryear}

\begin{document}

\begin{frontmatter}

\title{Solar Jet on 2014 April 16 Modeled by Kelvin--Helmholtz Instability}

\author{M.~Bogdanova}
\author{I.~Zhelyazkov\corref{mycorrespondingauthor}}
\address{Faculty of Physics, Sofia University, 1164 Sofia, Bulgaria}
\cortext[mycorrespondingauthor]{Corresponding author}
\ead{izh@phys.uni-sofia.bg}
\author{R.~Joshi, R.~Chandra}
\address{Department of Physics, DSB Campus, Kumaun University, Nainital 263001, India}

\begin{abstract}
We study here the arising of Kelvin--Helmholtz Instability (KHI) in one fast jet of 2014 April 16 observed by the Atmospheric Imaging Assembly (AIA) on board \emph{Solar Dynamics Observatory} (\emph{SDO}) in different UV and EUV wavelengths.  The evolution of jet indicates the blob like structure at its boundary which could be the observational evidence of the KHI.  We model the jet as a moving cylindrical magnetic flux tube of radius $a$ embedded in a magnetic field $\vec{B}_\mathrm{i}$ and surrounded by rest
magnetized plasma with magnetic field $\vec{B}_\mathrm{e}$.  We explore the propagation of the kink MHD mode along the jet that can become unstable against the KHI if its speed exceeds a critical value.  Concerning magnetic fields topology we consider three different configurations, notably of (i) spatially homogeneous magnetic fields (untwisted magnetic flux tube), (ii) internal (label `i') twisted magnetic field and external homogeneous one (label `e') (single-twisted flux tube), and (iii) both internal and external twisted magnetic fields (double-twisted magnetic flux tube).  Plasma densities in the two media $\rho_\mathrm{i}$ and $\rho_\mathrm{e}$ are assumed to be homogeneous.  The density contrast is defined in two ways: first as $\rho_\mathrm{e}/\rho_\mathrm{i}$ and second as $\rho_\mathrm{e}/(\rho_\mathrm{i} + \rho_\mathrm{e})$.  Computations show that the KHI can occur at accessible flow velocities in all the cases of untwisted and single-twisted flux tubes.  It turns out, however, that in the case of a double-twisted flux tube the KHI can merge at an accessible jet speed only when the density contrast is calculated from the ratio $\rho_\mathrm{e}/(\rho_\mathrm{i} + \rho_\mathrm{e})$.  Evaluated KHI developing times and kink mode wave phase velocities at wavelength of $4$~Mm lie in the ranges of $1$--$6.2$~min and $202$--$271$~km\,s$^{-1}$, respectively---all being reasonable for the modeled jet.

\end{abstract}

\begin{keyword}
\texttt{Sun: solar jets - Sun: reconnections - KH instability}
\end{keyword}

\end{frontmatter}

\section{Introduction}
\label{sec:intro}

Solar jets are small scale eruptions observed at different heights of the solar atmosphere. They are observed in the different
parts of the solar surface such as: in coronal hole regions \citep{Madjarska13,Zhelyazkov17}, active regions \citep{Schmieder13,Chandra15,Sterling17}, quiet regions (for example \citealp{Panesar16}).  Since their first observations in X-rays, a very remarkable progress have been made to understand the physics of the solar jets including the high resolution data of \emph{Solar Dynamics Observatory\/} (\emph{SDO}).

The solar jets can be divided into two categories, that is, ``standard'' and ``blow-out'' jets (\citealp{Moore13,Moore15,Raouafi16,Chandra17}).  Initially this classification was based on their morphology.
In the case of standard jets their spires are thin and narrow during their entire lifetime and their bases remained relatively dim, except for the commonly observed compact jet bright point (JBP) on one side of the jet's base.  In contrary to this the blow-out jets can be defined by broad spires.  A blow-out jet initiated as standard jet starts as a emerging bipole reconnecting
with ambient open field.  But sometime during this reconnection process, the emerging bipole is triggered unstable and erupts outward and it becomes the ``blow-out'' jet.  The ``blow-out'' jet is associated with a small flux rope eruption.  Sometime this erupting flux rope can become coronal mass ejection (CME) (for example see \citealp{Liu15,Chandra17}).  The interpretation of ``blow-out'' jets was first proposed by \cite{Moore10,Moore13}.  Later on, these were reported and confirmed in several observations \citep{Sterling16,Panesar16,Sterling17}.

The accepted mechanism for the generation of solar jets is the magnetic reconnection.  Conditions for the magnetic reconnection can be magnetic flux emergence \citep{Heyvaerts77,Shibata92,Guo13}, also flux cancellation \citep{Priest94,Longcope98,Innes10,Adams14}, or can be both \citep{Young14,Chandra15}. Several numerical simulations have been done in this direction.  Most of these includes the magnetic flux emergence. Moreover, in case of Pariat and co-workers \citep{Pariat17} one makes the condition for magnetic reconnection imposing horizontal photospheric twisting motions.
Very recently it is proposed that the jets are small scale phenomena and this physical mechanism is similar to the large scale eruptions \citep{Wyper17}.  Therefore to understand these small scale phenomena is very
crucial to explain the large scale solar eruptions.

Apart from this, the jets in the solar atmosphere can support the propagation of magnetohydrodynamic (MHD) waves, which becoming unstable can be one of the possible ways for the coronal heating.  Several models have been proposed for the heating of solar corona.  It is believed that there are two possible candidates for the coronal heating namely magnetic reconnection and the
dissipation of energy by MHD waves.  However, it is still not clear which one is the favorable mechanism.  Now people thinks it may be the combination of these two mechanisms.  It is known since long time that when the fluids of different speeds flow in the same direction, there will be a strong velocity shear near the interface region of two different speeds. This velocity shear
produced the vortex sheet at the boundaries.  As a result of this Kelvin--Helmholtz (KH) instability occurs.  The KHI MHD modeling have been done in several studies (for reviews see \cite{Zhelyazkov15, Nakariakov16}).  In addition to MHD modelings, there are few cases where the KHI vortex were observed \citep{Foullon11, Zhelyazkov15, Kuridze16, Zhelyazkov17}. The vortex sheet, which
is due to KHI can become unstable like spiral perturbations.  The vortex sheet causes the conversion of the directed flow energy into the turbulent energy (e.g., \cite{Maslowe85}).  And this energy can be one of the possible source of coronal heating.  Therefore, to study the KH instability in the solar features is very crucial and useful to understand the solar coronal heating
problem.  Keeping this in mind, we aim here the modeling of KH instability of one solar jet observed by the Atmospheric Imaging Assembly (AIA) on board \emph{SDO\/} on 2014 April 16.

The organization of the article is as follows: we present the observations and the selection criteria in Section 2. The jet and magnetic field geometry, as well as the governing MHD equations for the modeling are given in Section 3.  Section 4 deals with the derivation of wave dispersion relations obtained from our model.  Finally, in Sections 5 and 6, we discuss the numerical results and conclude our study.

\section{Observations}
\label{sec:observations}
During 2014 April 15--16, the active region NOAA AR 12035 produced several jets.  These jets are well observed by the Atmospheric Imaging Assembly (AIA) on board \emph{SDO\/} satellite.  The AIA observe the Sun in different UV and EUV wavelengths. The pixel size and temporal evolution of the AIA data is 0.6~arcsec and 12~s respectively.  The dynamics and the kinematics of these jets were described by \cite{Joshi17}. They found two jet sites close to each other and noticed the slippage of jets from one site to other site.  Using the Heliospheric Magnetic Imager (HMI) photospheric magnetic field data, they found that the jets site were
associated with the flux emergence as well the flux cancellation and presence of several null points (see their Figure 10).

In this article, we have selected one jet, notably J\ensuremath{'}${_6}$ from \citealp{Joshi17} Table 1 and model it for the Kelvin--Helmholtz instability.  Our selection criterion is based on its high speed, large size, longer life-time and visibility in all EUV channels relative to other jets.  In addition to this criterion, this jet indicates eruption of blobs type structures at its boundary. We believe that these blobs like structures could be the result of KHI.
Therefore, the chosen jet is a good candidate for KHI modeling.

The jet starts ${\approx}$14:47 UT and reached its maximum length {$\approx$}14:57 UT and finally finish around 15:00 UT.  The jet's maximum speed was ${\approx}343$~km\,s$^{-1}$ at AIA 211~\AA\ and the average speed of all the EUV channels was ${\approx}332$~km\,s$^{-1}$.

\begin{figure}[!h]
\centering
\includegraphics[width=1.0\textwidth,clip=]{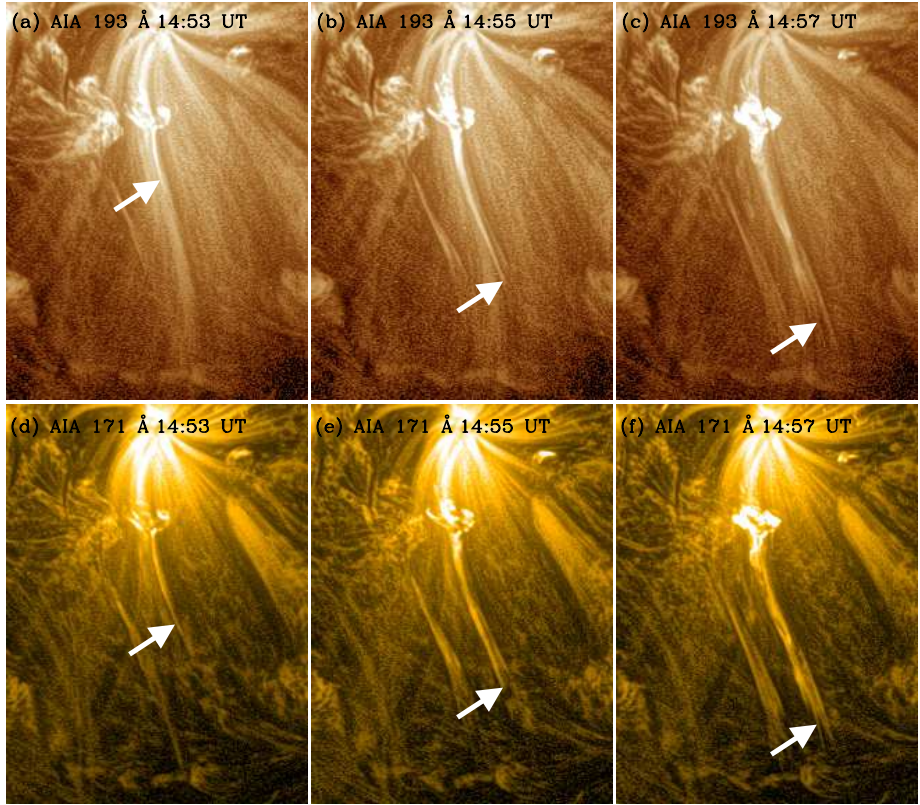}
\caption{Evolution of the jet in AIA 171 and 193~\AA.  The field-of-view is $130\times180$~arcsec.}
\label{fig:fig1}
\end{figure}
\begin{figure}[!h]
\centering
\includegraphics[width=0.45\textwidth,clip=]{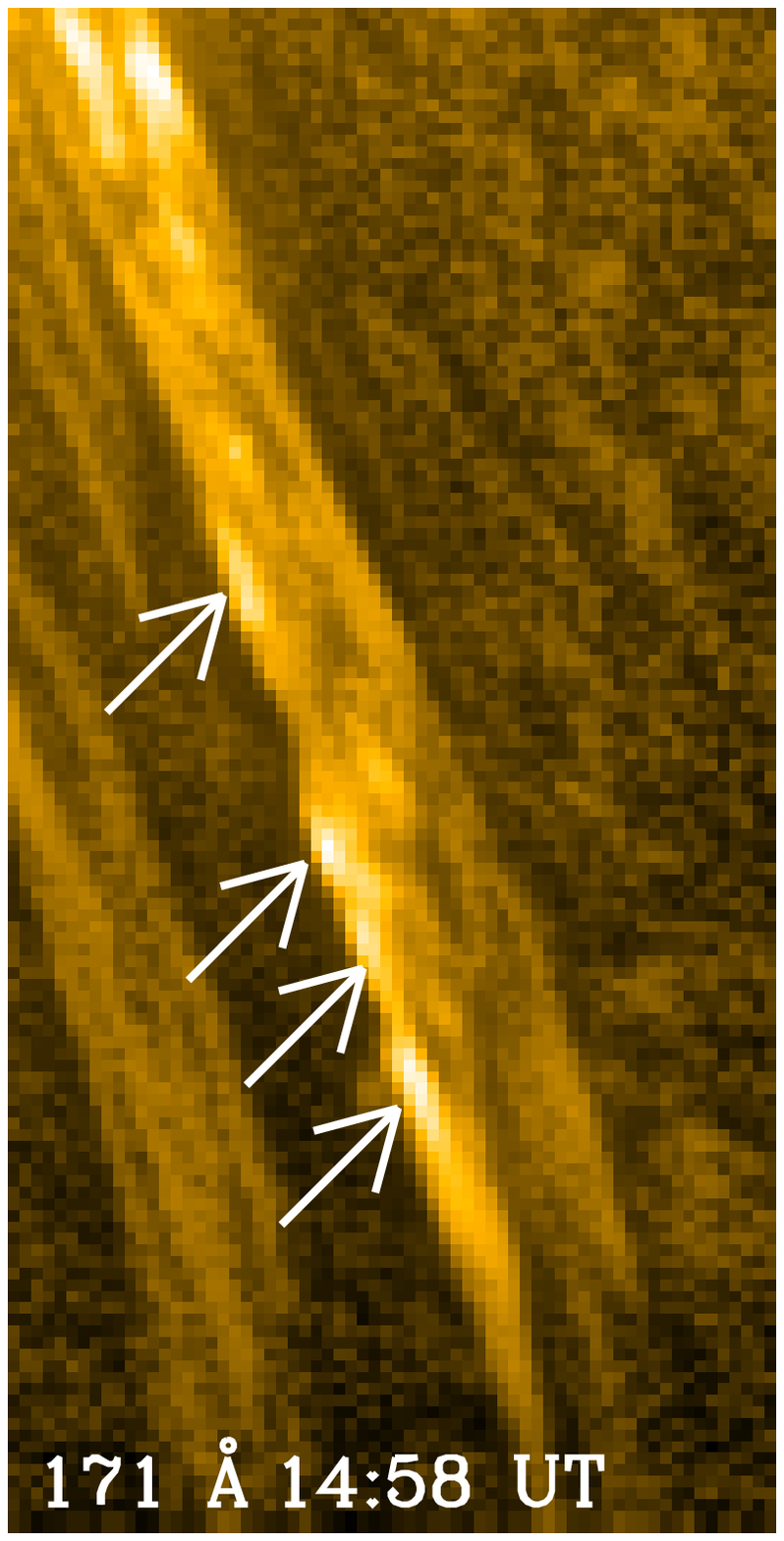}
\vspace*{-1cm}
\caption{AIA 171~\AA\ images of the jet.  The white arrows indicate different visible structures
of blobs, which could be due to KHI.  The field-of-view of the images is $40\times80$~arcsec.}
\label{fig:fig2}
\end{figure}
For a better image quality for \emph{SDO}/AIA data, we have used the Multi-Gaussian Normalization (MGN) techniques \citep{Morgan14} to enhanced the image quality.  The evolution of the jet in AIA/EUV 171 and 193~\AA\ is displayed in
Figure~\ref{fig:fig1}.  During jet's evolution, we have noticed the propagation of plasma blobs along the jet boundary in AIA 171~\AA.  To investigate these structures in more detail, we have enlarged the jet image in smaller field-of-view and this
is shown in Figure~\ref{fig:fig2}.  An inspection of these images evidenced the movement of the plasma blobs at the jet's boundary.  These structures are shown by the white arrows.  These blob structures evidenced the KHI.

\begin{figure}[!h]
\centering
\includegraphics[width=1.0\textwidth,clip=]{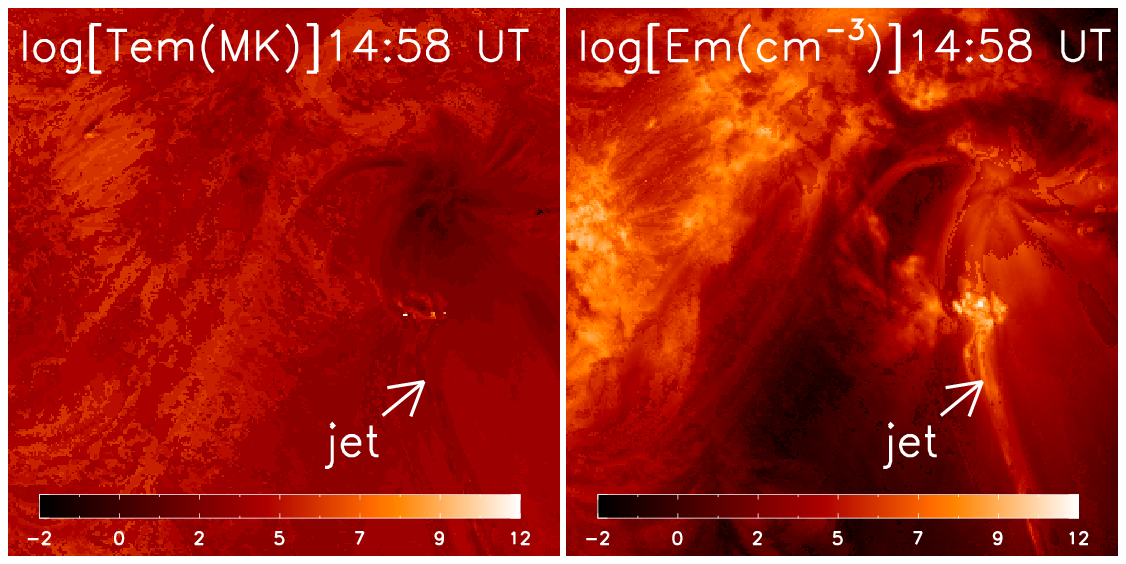}
\vspace*{-1cm}
\caption{Maps of temperature and emission measure inside and outside the jet derived by the \emph{SDO\/} data.}
\label{fig:fig3}
\end{figure}
To model the jet for the KHI, we have computed the temperature and the density inside and outside of the jet.  For this calculation, we have used the technique established by \cite{Aschwanden13}.  This techniques uses the observations of six AIA EUV channels, i.e., 94, 131, 171, 193, 211 and 335 \AA.  The estimated values for the jet are given in Table 1.  The subscripts `i' and `e' stamp for \emph{interior\/} and \emph{exterior}, respectively.  An example of temperature and emission measure maps are presented in Figure~\ref{fig:fig3} during the peak phase of the jet.
\begin{figure}[!t]
 \centerline{\includegraphics[width=0.65\textwidth,clip=]{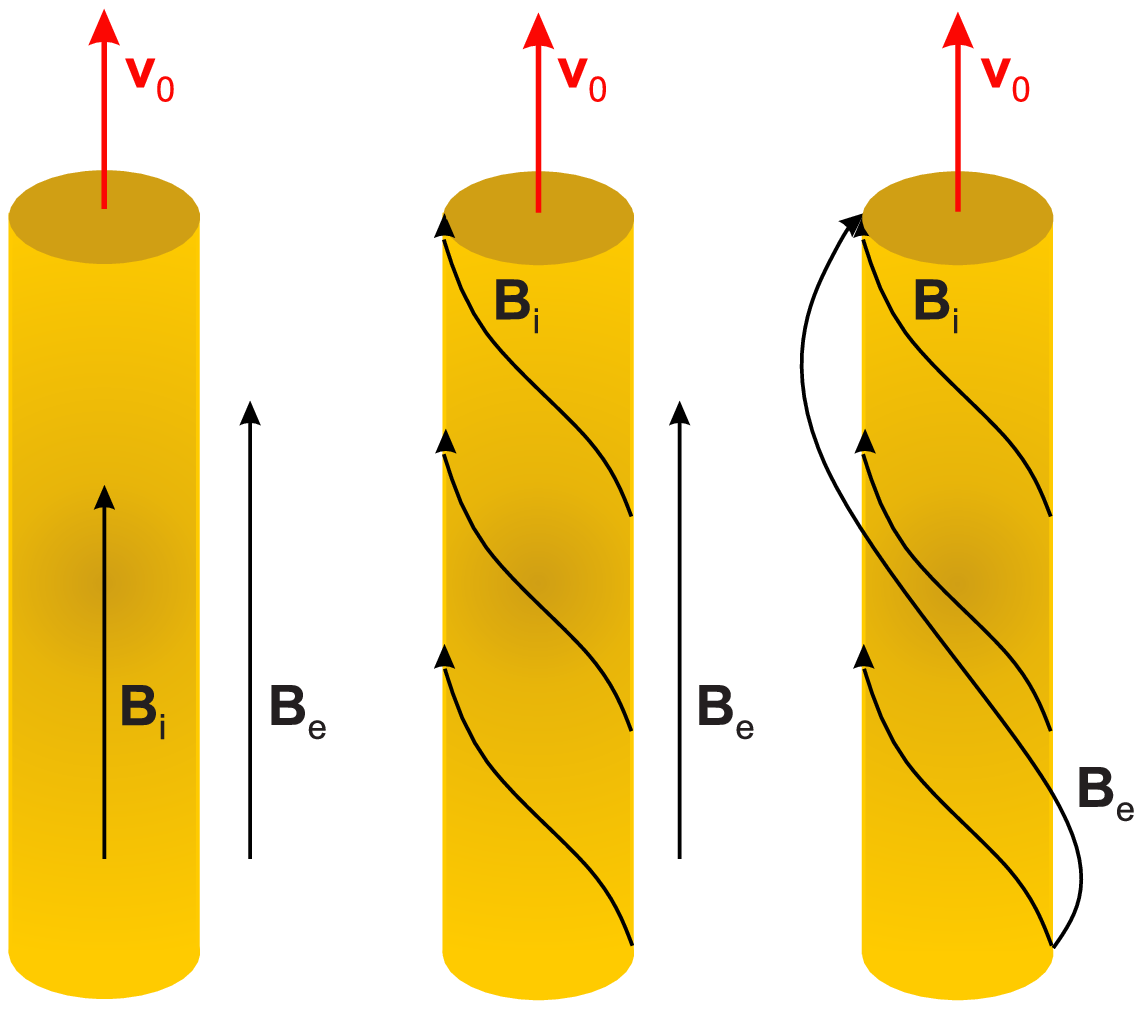}}
 \caption{Magnetic fields and flow velocity configurations in axially moving solar jet flux tubes.}
 \label{fig:fig4}
\end{figure}
%
%
\begin{table}
\caption{Different jet's physical parameters derived from \emph{SDO}/AIA data.}
\label{parameter}
\centering

\begin{tabular}{ccccc}
\hline
Maximum & Average speed & Temperature & Electron density\\
speed &  (EUV wavelengths) & (MK)  &  (${\times}10^9$ cm$^{-3}$)            \\
(km\,s$^{-1}$)  & (km\,s$^{-1}$) & $T_\mathrm{i}$~~~~~$T_\mathrm{e}$ & $\rho_\mathrm{i}$~~~~$\rho_\mathrm{e}$ \\
\hline
343 at 304 \AA\ & 332&1.61~~1.81 & 9.45~~6.38 \\
\hline
\end{tabular}
\end{table}
%


\section{Jet's geometry and the governing magnetohydrodynamic equations}
\label{sec:geometry}
We model the jet as a moving cylindrical magnetic flux tube of radius $a$ surrounded by a static coronal plasma.  We assume that the jet and ambient plasma densities, $\rho_\mathrm{i}$ and $\rho_\mathrm{e}$, respectively, are homogeneous.  Concerning magnetic field topology we consider three cases illustrated in Figure~\ref{fig:fig4}.  In the first case (the left column in the picture) the magnetic fields inside and outside the tube are homogeneous accordingly with magnitudes $B_\mathrm{i}$ and $B_\mathrm{e}$.  In the second case (the middle column in Figure~\ref{fig:fig4}) we assume that the magnetic field inside the jet/magnetic flux tube, $\boldsymbol{B}_\mathrm{i}$, is slightly twisted, while the one of the environment, $\boldsymbol{B}_\mathrm{e}$, is still homogeneous.  In the latter case (the right column in the picture) both magnetic fields are twisted.  The jet velocity $\boldsymbol{v}_0$ and wavevector $\boldsymbol{k}$ are directed along the central line of each flux tube, which (the line) is the $z$ axis of a cylindrical coordinate system that will be used in deriving the wave dispersion relations of propagating normal MHD modes in the jet.  The primary reason for considering three types of magnetic flux tubes is to see, first, how the magnetic field twist of the internal magnetic field $\boldsymbol{B}_\mathrm{i}$ will change/affect the condition for emerging of KHI of given MHD mode, and second, how an additionally involved twisted external magnetic field $\boldsymbol{B}_\mathrm{e}$ will change the picture.  In cylindrical coordinates all perturbations associated with the wave propagation are proportional to $\exp[\mathrm{i}(-\omega t + m\phi + k_z z)]$, where $\omega$ is the angular wave frequency, $m$ is the azimuthal mode number, and $k_z$ is the axial wavenumber, that is, $\boldsymbol{k} = (0, 0, k_z)$.  Among the various MHD modes the kink one with $m = 1$ turns out to be the most unstable against the KHI---that is why we will focus our study on this mode only.  The KHI occurs when the jet velocity $\boldsymbol{v}_0$ exceeds some critical value, $v_0^{\mathrm{cr}}$.  That critical flow velocity depends upon two important parameters of the jet--environment system, notably the density contrast between the two media and the ratio of external to internal magnetic fields (more correctly the ratio of their axial components) \citep{Zaqarashvili14}.  The density contrast, $\eta$, can be defined in two ways: (i) as the ratio $\rho_\mathrm{e}/\rho_\mathrm{i}$ (the standard definition), or (ii) as suggested by \cite{Paraschiv15}, as the ratio $\rho_\mathrm{e}/(\rho_\mathrm{i} + \rho_\mathrm{e})$.  For the magnetic fields ratio, $b$, we have, according to Figure~\ref{fig:fig4}, three definitions of $b$, namely equal to $B_\mathrm{e}/B_\mathrm{i}$ for a untwisted magnetic flux tube, $B_\mathrm{e}/B_{\mathrm{i}z}$ for a slightly twisted tube, and $B_{\mathrm{e}z}/B_{\mathrm{i}z}$ for the case when both magnetic field are twisted.

According to Table~1, the standard density contrast $\eta$ is equal to $0.675$.  The estimated temperatures inside and outside the jet define the following sound speeds in both media: $c_\mathrm{si} = 151$~km\,s$^{-1}$ and $c_\mathrm{se} = 158$~km\,s$^{-1}$, respectively. Regarding the coronal magnetic field value many people tried to find it through observations.  The methods include Helioseismology, Radio observations as well as magnetic field modeling.  According to very recent study done by \cite {Luna17} the coronal magnetic field value ranges from $7$ to $30$~G.  They have estimated these values using the Helioseismology of the filament and magnetic field modeling and the result from  both these methods are consistent.  Keeping these observational facts in mind, we assume that the background magnetic field is $B_\mathrm{e} = 7$~G, then the Alfv\'en speed in the environment, calculated from the standard formula $v_\mathrm{A} = B/\!\!\sqrt{\mu \rho}$, where $\mu$ being the vacuum permeability, is $v_\mathrm{Ae} = 191$~km\,s$^{-1}$.  The Alfv\'en speed inside the flux tube can be derived from the pressure balance equation (the sum of thermal and magnetic pressure to be constant, that is, to have the same values in both media) and it yields $v_\mathrm{Ai} = 132.5$~km\,s$^{-1}$.  This implies an internal magnetic field $B_\mathrm{i} = 5.9$~G; thus the magnetic fields ratio is $b = 1.18$.  Both plasma betas accordingly are $\beta_\mathrm{i} = 1.55$ and $\beta_\mathrm{e} = 0.83$.  Using the \cite{Paraschiv15} definition of the density contrast, we get a lower value for $\eta$, notably $0.403$.  The sound speeds are the same as listed above.  At this value of $\eta$, to satisfy the pressure balance equation, one needs to increase the background magnetic field: $B_\mathrm{e} = 8$~G fits the bill giving $v_\mathrm{Ae} = 218.3$~km\,s$^{-1}$, $v_\mathrm{Ai} = 63.2$~km\,s$^{-1}$, $b = 2.192$, $\beta_\mathrm{i} = 6.83$ and $\beta_\mathrm{e} = 0.63$, respectively.

With aforementioned density contrasts, sound and Alfv\'en speeds, as well as magnetic fields ratios, prior to starting numerical calculation of wave dispersion relations of the kink mode ($m = 1$), we can make some predictions first for the nature of the propagating normal mode (be it pure surface, pseudo surface/body, or leaky wave) that depends on the speeds ordering \citep{Cally86}; second, to evaluate the so called `kink speed $c_\mathrm{k}$' in a static magnetic flux tube expressed in terms of plasma densities and Alfv\'en speeds in both media \citep{Edwin83}:
\begin{equation}
\label{eq:kinkspeed}
    c_\mathrm{k} = \left( \frac{\rho_\mathrm{i} v_\mathrm{Ai}^2 + \rho_\mathrm{e}
        v_\mathrm{Ae}^2}{\rho_\mathrm{i} + \rho_\mathrm{e}} \right)^{1/2} = \left(
        \frac{1 + b^2}{1 + \eta} \right)^{1/2}v_\mathrm{Ai},
\end{equation}
which, as seen, is independent of sound speeds and characterizes the propagation of transverse perturbations; and third, to find the expected threshold/critical Alfv\'en Mach number $M_\mathrm{A}$ at which KHI would start--the later is determined by the inequality \citep{Zaqarashvili14}:
\begin{equation}
\label{eq:criterion}
    M_\mathrm{A}^2 > \left( 1 + 1/\eta \right)\left( b^2 + 1 \right).
\end{equation}
Here, the Alfv\'en Mach number is defined as $v_0/v_\mathrm{Ai}$.  These two formulas are written down for the kink mode propagating on a untwisted magnetic flux tube.  When the magnetic field $\boldsymbol{B}_\mathrm{i}$ is twisted, we should use modified values for the ratio $b$, and also for $v_\mathrm{Ai}$---for computational reasons, the Alfv\'en speed inside the jet should be defined as $B_{\mathrm{i}z}/\!\!\sqrt{\mu \rho_\mathrm{i}}$.

Dispersion relations of MHD modes are generally derived from the basic linearized equations of ideal magnetohydrodynamics.  Linearization implies that each variable (plasma density, fluid velocity, magnetic field, thermal and magnetic pressures) consists of an equilibrium value and its perturbation, and each term in the linearized MHD equations should contain only one perturbation.  We note the equilibrium quantities with subscript `$0$' and their perturbations with subscript `$1$'.  Dispersion relation of normal MHD modes in a moving untwisted magnetic flux tube are well known and their solution in complex variables is more or less straightforward.  With a twisted flux tube situation is more complicated.  Up to now we have no appropriate dispersion relation of normal MHD modes for compressible plasma magnetic flux tube.  That is why we are forced to do some approximations.  In the case $\eta = 0.675$ plasma betas, as we already computed, are $\beta_\mathrm{i} = 1.55$ and $\beta_\mathrm{e} = 0.83$, respectively, that is both are close to $1$.  Then, according to \cite{Zank93} both media can be considered as incompressible plasmas and the wave dispersion relation can be obtained from the linearized equation of the incompressible magnetohydrodynamics.  In the second case with $\eta = 0.403$ where $\beta_\mathrm{i} = 6.83$ and $\beta_\mathrm{e} = 0.63$, obviously the internal medium can be treated as incompressible plasma, but an adequate approximation for its environment would be a cool plasma.  As \cite{Zank93} claim, for the $\beta \ll 1$ regime, there is a strong tendency for nearly incompressible perturbations to propagate in a 1D direction parallel to the magnetic field.  Thus, in our study for KHI development of Alfv\'en-like perturbations/waves in the latter twisted solar jet it is appropriate to consider the jet as incompressible plasma and its environment as a cool, also incompressible, medium.

In summary, we have to obtain five different dispersion relations depending upon the nature of the jet medium and its environment, namely one for the case of untwisted flux tube when both plasmas are treated as compressible fluids; two equations for the case when the internal and external magnetic fields are twisted and the two media are considered as incompressible plasmas, and finally two equations for magnetically twisted flux tubes when the internal medium is assumed to be incompressible plasma while the external one is cool plasma.  The derivation of the MHD modes dispersion relation for an untwisted flux tube on using the basic equation of ideal magnetohydrodynamics is more or less straightforward and we will take its final form.  The same is the case when the internal magnetic field is twisted only.  Dispersion relation of MHD waves traveling on a double-twisted magnetic flux tube assuming incompressible plasma approximation in both media was obtained by \cite{Zaqarashvili14} on using the operator coefficient techniques developed by \cite{Goossens92}.  Here, we will re-derive that dispersion relation starting from the basic MHD equations governing the incompressible plasma dynamics.  Moreover, assuming a cool plasma environment, we will obtain a new form of the dispersion relation applicable in the case when internal plasma is incompressible and the surrounding medium is a cool plasma.  Fortunately, the governing equations for space and time evolution of fluid velocity $\vec{v}$, magnetic field $\vec{B}$, and pressure $p$ perturbations for incompressible and cool plasmas have practically the same form:
\begin{equation}
\label{eq:momentum}
    \rho_0 \frac{\partial}{\partial t}\boldsymbol{v}_1 + \rho_0 (\vec{v}_0 \cdot \nabla)\vec{v}_1 + \nabla \left( p_1 + \frac{\vec{B}_0 \cdot \vec{B}_1}{\mu} \right) - \frac{1}{\mu}(\vec{B}_0 \cdot \nabla)\vec{B}_1 - \frac{1}{\mu}(\vec{B}_1 \cdot \nabla)\vec{B}_0 = 0,
\end{equation}
\begin{equation}
\label{eq:faraday}
    \frac{\partial}{\partial t}\vec{B}_1 - \nabla (\vec{v}_1 \times \vec{B}_0) - \nabla (\vec{v}_0 \times \vec{B}_1) = 0,
\end{equation}
\begin{equation}
\label{eq:div_v}
    \nabla \cdot \vec{v}_1 = 0,
\end{equation}
\begin{equation}
\label{eq:div_B}
    \nabla \cdot \vec{B}_1 = 0.
\end{equation}
We note that in the case of cool plasma, the thermal pressure $p$ and its perturbation $p_1$ are equal to zero, and the fluid velocity perturbation, $\vec{v}_1$, in cylindrical coordinates, is presented as $\vec{v}_1 = (v_{1r},v_{1\phi}, 0)$, while the magnetic field perturbation has its three non-zero components, notably $\vec{B}_1 = (B_{1r}, B_{1\phi}, B_{1z})$.  Notice also that for a cool plasma the total pressure perturbation $p_\mathrm{tot} = p_1 + \vec{B}_0 \cdot \vec{B}_1/\mu$ reduces to magnetic pressure perturbation only.

\section{Dispersion Relations}
\label{sec:dispeqn}
\subsection{Dispersion relation in untwisted moving flux tube}
\label{subsec:untwisted}
We start with the simplest magnetic fields configuration pictured by the left column in Figure~\ref{fig:fig4}.  Dispersion relation of normal MHD modes propagating in a flowing compressible jet surrounded by a static compressible plasma reads \citep{Terra-Homem03,Nakariakov07,Zhelyazkov12}
\begin{eqnarray}
\label{eq:dispeq}
	\frac{\rho_\mathrm{e}}{\rho_\mathrm{i}}\left( \omega^2 - k_z^2 v_\mathrm{Ae}^2
        \right) \kappa_\mathrm{i}\frac{I_m^{\prime}(\kappa_\mathrm{i}a)}{I_m(\kappa_\mathrm{i}a)}
        - \left[ \left( \omega - \vec{k} \cdot
        \vec{v}_0 \right)^2 - k_z^2 v_\mathrm{Ai}^2 \right] \kappa_\mathrm{e}\frac{K_m^{\prime}(\kappa_\mathrm{e}a)}{K_m(\kappa_\mathrm{e}a)} = 0.
\end{eqnarray}
Here, $\kappa$ is the wave attenuation coefficient, which characterizes the space structure of the wave in the corresponding medium and whose squared magnitude is given by the expression
\begin{equation}
\label{eq:kappa}
	\kappa^2 = -\frac{\left[ \left( \omega - \vec{k} \cdot \vec{v}_0
        \right)^2 - k_z^2 c_{\rm s}^2 \right]\left[ \left( \omega - \vec{k}
        \cdot \vec{v}_0 \right)^2 - k_z^2 v_\mathrm{A}^2 \right]}{\left( c_\mathrm{
        s}^2 + v_\mathrm{A}^2 \right)\left[ \left( \omega - \vec{k} \cdot
        \vec{v}_0 \right)^2 - k_z^2 c_\mathrm{T}^2 \right]},
\end{equation}
where
\begin{equation}
\label{eq:c_T}
	c_\mathrm{T} = c_\mathrm{s} v_\mathrm{A}/\left( c_\mathrm{s}^2 + v_\mathrm{A}^2 \right)^{1/2}
\end{equation}
is the so-called \emph{tube velocity\/} \citep{Edwin83}.  As seen from the expressions for the attenuation coefficient and tube velocity, both quantities have different values inside and outside the flux magnetic tube owing to the different magnitudes of sound and Alfv\'en speeds in both media.  In dispersion equation~(\ref{eq:dispeq}), $I_m$ and $K_m$ are the modified Bessel functions, and the prime means differentiation with respect to the function argument.  It is worth noticing that the mode frequency inside the flux tube is Doppler-shifted.

\subsection{Dispersion relations in twisted moving flux tube}
\label{subsec:twisted}
If the magnetic field is twisted, we assume that in cylindrical coordinate system the magnetic field has the following form:
$\vec{B} = \left( 0, B_\phi(r), B_z(r) \right)$.  The unperturbed magnetic field and thermal pressure satisfy the pressure balance equation
\begin{equation}
\label{eq:pressurebal}
    \frac{\mathrm{d}}{\mathrm{d}r}\left( p + \frac{B_\phi^2 + B_z^2}{\mu} \right) = -\frac{B_\phi^2}{\mu r}.
\end{equation}
For simplicity we consider magnetic flux tube with uniform twist, that is
\begin{equation}
\label{eq:B_i}
    \vec{B}_\mathrm{i} = (0, Ar, B_{\mathrm{i}z}),
\end{equation}
where $A$ and $B_{\mathrm{i}z}$ are constants.  The magnetic field in the environment is homogeneous, that is, $\vec{B}_\mathrm{e} = (0, 0, B_\mathrm{e})$.  The flow velocity of the moving flux tube is also homogeneous: $\vec{v}_0 = (0, 0, v_0)$.  When considering that both media are incompressible plasmas, from aforementioned set of equations (\ref{eq:momentum})--(\ref{eq:div_B}) and appropriate boundary conditions for continuity of total Lagrangian pressure perturbation and the transfer Lagrangian displacement $\xi_r$ (obtainable from the relation $v_{1r} = \partial \xi_r/\partial t$), one can get the following dispersion relation \citep{Zhelyazkov12a}:
\begin{equation}
\label{eq:twdispeq}
	\frac{\left( \Omega^2 -
    \omega_\mathrm{Ai}^2 \right)F_m(\kappa_\mathrm{i}a) - 2mA \omega_\mathrm{Ai}/\sqrt{\mu \rho_\mathrm{i}}}
    {\left( \Omega^2 -
    \omega_\mathrm{Ai}^2 \right)^2 - 4A^2\omega_\mathrm{Ai}^2/\mu \rho_\mathrm{i} }
    = \frac{P_m(k_z a)}
    { \left(\rho_\mathrm{e}/\rho_\mathrm{i}\right) \left( \omega^2 - \omega_\mathrm{Ae}^2
    \right) + A^2 P_m(k_z a)/\mu \rho_\mathrm{i}},
\end{equation}
where $\Omega = \omega - \vec{k}\cdot \vec{v}_0$ is the Doppler-shifted wave frequency in the moving medium,
\[
    F_m(\kappa_\mathrm{i}a) = \frac{\kappa_\mathrm{i}a I_m^{\prime}(\kappa_\mathrm{i}a)}{I_m(\kappa_\mathrm{i}a)}, \qquad
    P_m(k_z a) = \frac{k_z a K_m^{\prime}(k_z a)}{K_m(k_z a)},
\]
\[
    \omega_\mathrm{Ai} = \frac{\vec{k}\cdot \vec{B}_\mathrm{i}}{\sqrt{\mu \rho_\mathrm{i}}} = \frac{1}{\sqrt{\mu \rho_\mathrm{i}}} \left( \frac{m}{r}B_{\mathrm{i}\phi} + k_z B_{\mathrm{i}z} \right) \quad \mbox{and} \quad \omega_\mathrm{Ae} = \frac{k_z B_\mathrm{e}}{\sqrt{\mu \rho_\mathrm{e}}}
\]
are the local Alfv\'en frequencies inside the moving flux tube and outside its static environment.  It is easy to see that $\omega_\mathrm{Ae} = k_z v_\mathrm{Ae}$.  The wave attenuation coefficient inside the tube, $\kappa_\mathrm{i}$, is given by
\begin{equation}
\label{eq:kappa_i}
    \kappa_\mathrm{i} = k_z \left[ 1 - 4A^2 \omega_\mathrm{Ai}^2/\mu \rho_\mathrm{i} \left( \Omega^2 - \omega_\mathrm{Ai}^2 \right)^2 \right]^{1/2},
\end{equation}
while that in the surrounding medium, $\kappa_\mathrm{e}$, is simply equal to $k_z$.

In the case when the environment is treated as a cool medium, the form of the wave dispersion relation is the same as in Equation~(\ref{eq:twdispeq}), but the wave attenuation coefficient $\kappa_\mathrm{e} = k_z$ should be replaced by
\begin{equation}
\label{eq:kappa_e}
    \kappa_\mathrm{e} = \left( k_z^2 v_\mathrm{Ae}^2 - \omega^2 \right)^{1/2}v_\mathrm{Ae}.
\end{equation}

\subsection{Dispersion relations in twisted moving flux tube and twisted external magnetic field}
\label{subsec:dbletwisted}
The most complicated case is that pictured by the right column in Figure~\ref{fig:fig4}.  The basic idea in deriving the wave dispersion relation is to find out solutions for the total pressure perturbation, $p_\mathrm{tot}$, and perturbed interface, $\xi_r$, and merge them via appropriate boundary conditions.  The equilibrium magnetic field inside the moving flux tube is the same as in the previous subsection~\ref{subsec:twisted}, namely $\vec{B}_\mathrm{i} = (0, Ar, B_{\mathrm{i}z})$, where, recall, $A$ and $B_{\mathrm{i}z}$ are constants.  The jet's flow velocity has only $z$ component, equal to $v_0$.
As unperturbed parameters depend on the $r$ coordinate only, all the perturbations can be Fourier analyzed with $\exp[\mathrm{i}(-\omega t + m\phi + k_z z)]$ and the governing MHD equations (\ref{eq:momentum}) and (\ref{eq:faraday}) are:
\begin{equation}
\label{eq:v1r}
    -\mathrm{i}(\omega - \vec{k}\cdot \vec{v}_0)v_{1r} + \frac{1}{\rho_\mathrm{i}} \frac{\mathrm{d}}{\mathrm{d}r} p_\mathrm{tot} - \mathrm{i}\frac{1}{\sqrt{\mu \rho_\mathrm{i}}}\omega_\mathrm{Ai}B_{1r} + \frac{2A}{\mu \rho_\mathrm{i}}B_{1\phi} = 0,
\end{equation}
\begin{equation}
\label{eq:v1phi}
    -\mathrm{i}(\omega - \vec{k}\cdot \vec{v}_0)v_{1\phi} + \mathrm{i}\frac{1}{\rho_\mathrm{i}}\frac{m}{r}p_\mathrm{tot} - \mathrm{i}\frac{1}{\sqrt{\mu \rho_\mathrm{i}}}\omega_\mathrm{Ai}B_{1\phi} - \frac{2A}{\mu \rho_\mathrm{i}}B_{1r} = 0,
\end{equation}
\begin{equation}
\label{eq:v1z}
    -(\omega - \vec{k}\cdot \vec{v}_0)v_{1z} + \frac{1}{\rho_\mathrm{i}}k_z p_\mathrm{tot} - \frac{1}{\sqrt{\mu \rho_\mathrm{i}}}\omega_\mathrm{Ai}B_{1z} = 0,
\end{equation}
\begin{equation}
\label{eq:B1r}
    (\omega - \vec{k}\cdot \vec{v}_0)B_{1r} + (mA + k_z B_{\mathrm{i}z})v_{1r} = 0,
\end{equation}
\begin{equation}
\label{eq:B1phi}
    (\omega - \vec{k}\cdot \vec{v}_0)B_{1\phi} + (mA + k_z B_{\mathrm{i}z})v_{1\phi} = 0,
\end{equation}
\begin{equation}
\label{eq:B1z}
    (\omega - \vec{k}\cdot \vec{v}_0)B_{1z} + (mA + k_z B_{\mathrm{i}z})v_{1z} = 0,
\end{equation}
where
\begin{equation}
\label{eq:omegaAi}
    \omega_\mathrm{Ai} = \frac{1}{\sqrt{\mu \rho_\mathrm{i}}}\left( mA + k_z B_{\mathrm{i}z} \right)
\end{equation}
is the local Alfv\'en frequency inside the moving flux tube.  From the above equations one can obtain expressions for the fluid velocity components in terms of the total pressure perturbation, $p_\mathrm{tot}$, and its derivative with respect to $r$, namely
\begin{equation}
\label{eq:v1rnew}
    v_{1r} = -\mathrm{i}\frac{1}{Y_\mathrm{i}}\frac{1}{\Omega_\mathrm{i}\rho_\mathrm{i}}\left( \frac{\mathrm{d}}{\mathrm{d}r} - Z_\mathrm{i} \frac{m}{r} \right)p_\mathrm{tot},
\end{equation}
\begin{equation}
\label{eq:v1phinew}
    v_{1\phi} = \frac{1}{Y_\mathrm{i}}\frac{1}{\Omega_\mathrm{i}\rho_\mathrm{i}}\left( \frac{m}{r} - Z_\mathrm{i} \frac{\mathrm{d}}{\mathrm{d}r} \right)p_\mathrm{tot},
\end{equation}
\begin{equation}
\label{eq:v1znew}
    v_{1z} = \frac{1}{\Omega_\mathrm{i}\rho_\mathrm{i}}k_z p_\mathrm{tot}.
\end{equation}
Here,
\[
    \frac{1}{\Omega_\mathrm{i}} \equiv \frac{\omega - \vec{k}\cdot \vec{v}_0}{(\omega - \vec{k}\cdot \vec{v}_0)^2 - \omega_\mathrm{Ai}^2}, \quad Z_\mathrm{i} \equiv \frac{2A\omega_\mathrm{Ai}}{\sqrt{\mu \rho_\mathrm{i}}\left[ (\omega - \vec{k}\cdot \vec{v}_0)^2 - \omega_\mathrm{Ai}^2 \right]}, \quad \mbox{and} \quad Y_\mathrm{i} = 1 - Z_\mathrm{i}^2.
\]
On using these expressions for fluid velocity perturbation in the constraint equation (\ref{eq:div_v}), after some algebra we get that the total pressure perturbation obeys the equation
\begin{equation}
\label{eq:BesselfuncI}
    \left[ \frac{\mathrm{d}^2}{\mathrm{d}r^2} + \frac{1}{r}\frac{\mathrm{d}}{\mathrm{d}r} - \left( \frac{m^2}{r^2} + \kappa_\mathrm{i}^2\right) \right]p_\mathrm{tot} = 0,
\end{equation}
where $\kappa_\mathrm{i}$, not surprisingly, coincides with expression (\ref{eq:kappa_i}), that is,
\[
    \kappa_\mathrm{i}^2 = k_z^2 \left[ 1 - \frac{4A^2 \omega_\mathrm{Ai}^2}{\mu \rho_\mathrm{i}\left( \Omega^2 - \omega_\mathrm{Ai}^2 \right)} \right], \quad \mbox{where as before} \quad \Omega \equiv \omega - \vec{k}\cdot \vec{v}_0.
\]

Equation~(\ref{eq:BesselfuncI}) is the Bessel equation whose solution bounded at the tube axis is
\begin{equation}
\label{eq:ptoti}
    p_\mathrm{i\,tot} = \alpha_\mathrm{i}I_m(\kappa_\mathrm{i}r).
\end{equation}
Here, $I_m$ is the modified Bessel function of order $m$ and $\alpha_\mathrm{i}$ is a constant.  Transverse displacement, $\xi_{\mathrm{i}r}$, can be obtained from expression~(\ref{eq:v1rnew}) and has the form:
\begin{equation}
\label{eq:xi_i}
    \xi_{\mathrm{i}r} = \frac{\alpha_\mathrm{i}}{r} \frac{\left( \Omega^2 - \omega_\mathrm{Ai}^2 \right)\kappa_\mathrm{i}rI_m^{\prime}(\kappa_\mathrm{i}r) - 2mA\omega_\mathrm{Ai} I_m(\kappa_\mathrm{i}r)/\!\!\sqrt{\mu \rho_\mathrm{i}}}{\rho_\mathrm{i}\left( \Omega^2 - \omega_\mathrm{Ai}^2 \right)^2 - 4A^2 \omega_\mathrm{Ai}^2/\mu},
\end{equation}
where the prime, $^{\prime}$, implies a differentiation by the Bessel function argument.

Let us now go to the environment and find similar expressions for $p_\mathrm{e\,tot}$ and $\xi_{\mathrm{e}r}$, respectively.
We consider the external magnetic field of the form
\begin{equation}
\label{eq:B_external}
    \vec{B}_\mathrm{e} = \left( 0, B_{\mathrm{e}\phi}(a)\frac{a}{r}, B_{\mathrm{e}z}(a)\frac{a^2}{r^2} \right)
\end{equation}
(where for convenience we denote $B_{\mathrm{e}\phi}(a) \equiv B_\phi$ and $B_{\mathrm{e}z}(a) \equiv B_z$) and the plasma density in the form $\rho_0 = \rho_\mathrm{e}(a/r)^4$ so that the Alfv\'en frequency
\begin{eqnarray}
\label{eq:omegaAe}
    \omega_\mathrm{Ae} = \frac{1}{\sqrt{\mu \rho_0(r)}}\left[ \frac{m}{r}B_{\mathrm{e}\phi}(r) + k_z B_{\mathrm{e}z}(r) \right] = \frac{r^2}{\sqrt{\mu \rho_\mathrm{e}a^4}}\left( \frac{maB_\phi}{r^2} + \frac{k_z a^2 B_z}{r^2} \right) \nonumber \\
    \nonumber \\
    {}= \frac{mB_\phi + k_z aB_z}{\sqrt{\mu \rho_\mathrm{e} a^2}}
\end{eqnarray}
is constant.  This circumstance allows us to find analytical solution to the governing equations.  Now momentum and Faraday equations are displayed in the form:
\begin{equation}
\label{eq:v1rext}
    -\mathrm{i}\omega v_{1r} + \frac{1}{\rho_\mathrm{e}}\frac{\mathrm{d}}{\mathrm{d}r}p_\mathrm{tot} - \mathrm{i}\frac{1}{\mu \rho_\mathrm{e}}\left( \frac{maB_\phi}{r^2} + \frac{k_z a^2 B_z}{r^2} \right)B_{1r} + 2\frac{a}{r^2}
    \frac{B_\phi}{\mu \rho_\mathrm{e}}B_{1\phi} = 0,
\end{equation}
\begin{equation}
\label{eq:v1phiext}
    -\mathrm{i}\omega v_{1\phi} + \frac{1}{\rho_\mathrm{e}}\frac{m}{r}p_\mathrm{tot} - \mathrm{i}\frac{1}{\mu \rho_\mathrm{e}}\left( \frac{maB_\phi}{r^2} + \frac{k_z a^2 B_z}{r^2} \right)B_{1\phi} = 0,
\end{equation}
\begin{equation}
\label{eq:v1zext}
    -\mathrm{i}\omega v_{1z} + \frac{1}{\rho_\mathrm{e}}k_z p_\mathrm{tot} - \mathrm{i}\frac{1}{\mu \rho_\mathrm{e}}\left( \frac{maB_\phi}{r^2} + \frac{k_z a^2 B_z}{r^2} \right)B_{1z} + 2\frac{a^2}{r^3}
    \frac{B_z}{\mu \rho_\mathrm{e}}B_{1r} = 0,
\end{equation}
\begin{equation}
\label{eq:B1rext}
    \omega B_{1r} + \left( \frac{maB_\phi}{r^2} + \frac{k_z a^2 B_z}{r^2} \right)v_{1r} = 0,
\end{equation}
\begin{equation}
\label{eq:B1phiext}
    \mathrm{i}\omega B_{1\phi} + \mathrm{i}\left( \frac{maB_\phi}{r^2} + \frac{k_z a^2 B_z}{r^2} \right)v_{1\phi} + 2 \frac{a}{r^2}B_\phi v_{1r} = 0,
\end{equation}
\begin{equation}
\label{eq:B1zext}
    \mathrm{i}\omega B_{1z} + \mathrm{i}\left( \frac{maB_\phi}{r^2} + \frac{k_z a^2 B_z}{r^2} \right)v_{1z} + 2 \frac{a^2}{r^3}B_z v_{1r} = 0.
\end{equation}

By using expression~(\ref{eq:omegaAe}), from above equations, after a lengthy algebra one obtains that
\begin{equation}
\label{eq:v1rextnew}
    v_{1r} = -\mathrm{i}\frac{1}{Y_\mathrm{e}}\frac{1}{\Omega_\mathrm{e}\rho_\mathrm{e}}\left( \frac{\mathrm{d}}{\mathrm{d}r} - Z_\mathrm{e}\frac{m}{r} \right)p_\mathrm{tot},
\end{equation}
\begin{equation}
\label{eq:v1phiextnew}
    v_{1\phi} = \frac{1}{\Omega_\mathrm{e}\rho_\mathrm{e}}\left[ \left( 1 + \frac{Z_\mathrm{e}^2}{Y_\mathrm{e}} \right) \frac{m}{r} - \frac{Z_\mathrm{e}}{Y_\mathrm{e}} \frac{\mathrm{d}}{\mathrm{d}r} \right]p_\mathrm{tot},
\end{equation}
\begin{equation}
\label{eq:v1zextnew}
    v_{1z} = \frac{1}{\Omega_\mathrm{e}\rho_\mathrm{e}}k_z p_\mathrm{tot},
\end{equation}
where now
\[
    \frac{1}{\Omega_\mathrm{e}} \equiv \frac{\omega}{\omega^2 - \omega_\mathrm{Ae}^2}, \quad Z_\mathrm{e} \equiv \frac{2B_\phi \omega_\mathrm{Ae}}{\sqrt{\mu \rho_\mathrm{e}a^2}\left(\omega^2 - \omega_\mathrm{Ae}^2 \right)}, \quad \mbox{and} \quad Y_\mathrm{e} = 1 - \frac{\omega^2}{\omega_\mathrm{Ae}^2}Z_\mathrm{e}^2.
\]

With these expressions for fluid velocity perturbations, on using again the equation $\nabla \cdot \vec{v}_1 = 0$, we obtain that the total pressure perturbation obeys the equation
\begin{equation}
\label{eq:BesselfuncK_nu}
    \left[ \frac{\mathrm{d}^2}{\mathrm{d}r^2} + \frac{5}{r}\frac{\mathrm{d}}{\mathrm{d}r} - \left( \frac{n^2}{r^2} + \kappa_\mathrm{e}^2\right) \right]p_\mathrm{tot} = 0,
\end{equation}
which is an equation for Bessel function with complex order $\nu = \sqrt{4 + n^2}$, with a solution bounded at infinity in the form
\begin{equation}
\label{eq:ptote}
    p_\mathrm{e\,tot} = \alpha_\mathrm{e} \frac{a^2}{r^2} K_\nu(\kappa_\mathrm{e}r),
\end{equation}
where $\alpha_\mathrm{e}$ is a constant, the wave attenuation coefficient is equal to
\begin{equation}
\label{eq:kappa_eext}
    \kappa_\mathrm{e} = k_z \left( 1 - \frac{4B_\phi^2 \omega^2}{\mu \rho_\mathrm{e} \left( \omega^2 - \omega_\mathrm{Ae}^2 \right)^2 a^2} \right)^{1/2},
\end{equation}
and the term $n^2$ is given by the expression
\[
    n^2 = m^2 - \frac{4m^2 B_\phi^2}{\mu \rho_\mathrm{e}a^2 \left( \omega^2 - \omega_\mathrm{Ae}^2 \right)} + \frac{8mB_\phi \omega_\mathrm{Ae}}{\sqrt{\mu \rho_\mathrm{e}}a \left( \omega^2 - \omega_\mathrm{Ae}^2 \right)}.
\]

With the help of the relation $\xi_r = \mathrm{i}v_{1r}/\omega$, on using Equation~(\ref{eq:v1rextnew}), one obtains that
\begin{eqnarray}
\label{eq:xi_e}
    \xi_{\mathrm{e}r} = \alpha_\mathrm{e} \frac{r\left( \omega^2 - \omega_\mathrm{Ae}^2 \right)\kappa_\mathrm{e}r K_\nu^{\prime}(\kappa_\mathrm{e}r)}{a^2 \rho_\mathrm{e}\left( \omega^2 - \omega_\mathrm{Ae}^2 \right)^2 - 4B_\phi^2 \omega^2/\mu} \nonumber \\
    \nonumber \\
    {}- \alpha_\mathrm{e} \frac{r}{a} \left( \frac{2a \left( \omega^2 - \omega_\mathrm{Ae}^2 \right) + 2mB_\phi \omega_\mathrm{Ae}/\!\!\sqrt{\mu \rho_\mathrm{e}}}{a^2 \rho_\mathrm{e}\left( \omega^2 - \omega_\mathrm{Ae}^2 \right)^2 - 4B_\phi^2 \omega^2/\mu} \right) K_\nu(\kappa_\mathrm{e}r).
\end{eqnarray}
Merging the solutions for $p_\mathrm{tot}$ and $\xi_r$ in both media at the tube surface, $r = a$, one obtains the dispersion equation of the normal MHD modes propagating on the moving magnetic flux tube.  As we already said, the boundary conditions at the tube surface are the continuity of the Lagrangian displacement
\begin{equation}
\label{eq:boundarycond1}
    \xi_{\mathrm{i}r}|_{r=a} = \xi_{\mathrm{e}r}|_{r=a}
\end{equation}
(where $\xi_{\mathrm{i}r}$ and $\xi_{\mathrm{e}r}$ are given by Equations~(\ref{eq:xi_i}) and (\ref{eq:xi_e})), and the total Lagrangian pressure \citep{Bennett99}
\begin{equation}
\label{eq:boundarycond2}
        \left.p_\mathrm{i\,tot} - \frac{B_{\mathrm{i}\phi}^2}{\mu a}\xi_{\mathrm{i}r}\right\vert_{r=a} = \left.p_\mathrm{e\,tot} - \frac{B_{\mathrm{e}\phi}^2}{\mu a}\xi_{\mathrm{e}r}\right\vert_{r=a},
\end{equation}
where the total pressure perturbations are given by Equations~(\ref{eq:ptoti}) and (\ref{eq:ptote}), respectively.  Using these boundary conditions we recover the dispersion equation governing the oscillations in moving twisted flux tube surrounded by twisted incompressible magnetized plasma derived in \citealp{Zaqarashvili14}:
\begin{eqnarray}
\label{eq:dbletwdispeq}
	\frac{\left( \left[ \omega - \vec{k} \cdot \vec{v}_0 \right]^2 -
    \omega_\mathrm{Ai}^2 \right)F_m(\kappa_\mathrm{i}a) - 2mA \omega_\mathrm{Ai}/\sqrt{\mu \rho_\mathrm{i}}}
    {\rho_\mathrm{i}\left( \left[ \omega - \vec{k} \cdot \vec{v}_0 \right]^2 -
    \omega_\mathrm{Ai}^2 \right)^2 - 4A^2\omega_\mathrm{Ai}^2/\mu} \nonumber \\
    \nonumber \\
    {}= \frac{a^2 \left( \omega^2 - \omega_\mathrm{Ae}^2 \right)Q_\nu(\kappa_\mathrm{e}a) - G}
    {L - H \left[a^2 \left( \omega^2 - \omega_\mathrm{Ae}^2 \right)Q_\nu(\kappa_\mathrm{e}a) - G \right]},
\end{eqnarray}
where
\[
    F_m(\kappa_\mathrm{i}a) = \frac{\kappa_\mathrm{i}a I_m^{\prime}(\kappa_\mathrm{i}a)}{I_m(\kappa_\mathrm{i}a)}, \quad
    Q_\nu(\kappa_\mathrm{e}a) = \frac{\kappa_\mathrm{e}a K_\nu^{\prime}(\kappa_\mathrm{e}a)}{K_\nu(\kappa_\mathrm{e}a)}, \quad L = a^2 \rho_\mathrm{e} \left( \omega^2 - \omega_\mathrm{Ae}^2 \right)^2 - 4B_{\mathrm{e}\phi}^2 \omega^2/\mu,
\]
\[
     H = B_{\mathrm{e}\phi}^2/\mu a^2 - A^2/\mu, \quad G = 2a^2 \left( \omega^2 - \omega_\mathrm{Ae}^2 \right) + 2maB_{\mathrm{e}\phi} \omega_\mathrm{Ae}/\!\!\sqrt{\mu \rho_\mathrm{e}}.
\]

When surrounding plasma is treated as a cool medium, then, recall, the plasma pressure perturbation $p_1 = 0$ and the total pressure perturbation, $p_\mathrm{tot}$ reduces to magnetic pressure perturbation only.  Moreover, the axial component of fluid velocity perturbation also is zero, that is, $v_{1z} = 0$.  Under these circumstances, among the six governing MHD equations (\ref{eq:v1rext})--(\ref{eq:B1zext}), only Equations~(\ref{eq:v1zext}) and (\ref{eq:B1zext}) are slightly changed; we will denote the magnetic pressure perturbation by $p_\mathrm{mag1}$.  By using Equation~(\ref{eq:div_B}), we express $B_{1z}$ in terms of $B_{1r}$ and $B_{1\phi}$ and obtain
\[
    B_{1z} = \mathrm{i}\frac{1}{k_z}\left( \frac{\mathrm{d}}{\mathrm{d}r} + \frac{1}{r} \right)B_{1r} - \frac{1}{k_z}\frac{m}{r}B_{1\phi}.
\]
Inserting this $B_{1z}$ alongside with $B_{1r}$ and $B_{1\phi}$ expressed via $v_{1r}$ and $v_{1\phi}$ (by using Equations~(\ref{eq:B1rext}) and (\ref{eq:B1phiext})) into Equation~(\ref{eq:v1zext}), we get
\[
    -\mathrm{i}\frac{1}{\rho_\mathrm{i}}k_z^2 p_\mathrm{mag1} - \frac{\omega_\mathrm{Ae}^2}{\omega}\left( \frac{\mathrm{d}}{\mathrm{d}r} + 3\frac{1}{r} \right)v_{1r} - \mathrm{i}\frac{\omega_\mathrm{Ae}^2}{\omega} \frac{m}{r} v_{1\phi} = 0.
\]
Now, on using Equations~(\ref{eq:v1rextnew}) and (\ref{eq:v1phiextnew}) giving us fluid velocity perturbation components in terms of total (that is, magnetic) pressure perturbation, above equation takes the form
\begin{equation}
\label{eq:BesselfuncK_nucool}
    \left[ \frac{\mathrm{d}^2}{\mathrm{d}r^2} + \frac{3}{r}\frac{\mathrm{d}}{\mathrm{d}r} - \left( \frac{n_\mathrm{c}^2}{r^2} + \kappa_\mathrm{ec}^2\right) \right]p_\mathrm{mag1} = 0,
\end{equation}
which like before is the equation for Bessel function with complex order $\nu_\mathrm{c} = \sqrt{2 + n_\mathrm{c}^2}$, where
\[
    n_\mathrm{c}^2 = m^2 - \frac{4m^2 B_\phi^2}{\mu \rho_\mathrm{e}a^2 \left( \omega^2 - \omega_\mathrm{Ae}^2 \right)} + \frac{4mB_\phi \omega_\mathrm{Ae}}{\sqrt{\mu \rho_\mathrm{e}}a \left( \omega^2 - \omega_\mathrm{Ae}^2 \right)}.
\]
Here, the label `c' stamps for \emph{cool}.  Note that in the cool environment the wave attenuation coefficient is given by
\begin{equation}
\label{eq:kappa_eextc}
    \kappa_\mathrm{ec} = k_z \left( 1 - \frac{4B_\phi^2 \omega^2}{\mu \rho_\mathrm{e} \left( \omega^2 - \omega_\mathrm{Ae}^2 \right)^2 a^2} \right)^{1/2}\left( \frac{\omega^2}{\omega_\mathrm{Ae}^2} - 1 \right)^{1/2}.
\end{equation}
Further on, following the standard steps for deriving the wave dispersion relation, we arrive at Equation~(\ref{eq:dbletwdispeq}) in which have to replace $\nu$ by $\nu_\mathrm{c}$ and $\kappa_\mathrm{e}$ by $\kappa_\mathrm{ec}$, respectively.

Having derived all necessary wave dispersion equations, we can now apply them in studying the propagation characteristics primarily of the kink ($m = 1$) mode running on the observed EUV jet, that will be done in the next section.

\section{Numerical results and discussion}
\label{sec:numerics}
Since we are looking for unstable solutions to wave dispersion relations, assume the MHD modes are running along the flux tube having a real axial wavenumber $k_z$ and a complex angular wave frequency $\omega \equiv \mathrm{Re}\omega + \mathrm{i}\,\mathrm{Im}\omega$.  Numerical results are usually presented as dependence of the complex wave phase velocity $v_\mathrm{ph} = \omega/k_z$ on the $k_z$.  For convenience, we normalize all velocities with respect to the Alfv\'en speed inside the tube, $v_\mathrm{Ai}$, and the wavelength $\lambda = 2\pi/k_z$ with respect to the tube radius, $a$.  Thus we have a complex dimensionless wave phase velocity $v_\mathrm{ph}/v_\mathrm{Ai}$ and dimensionless axial wavenumber $k_z a$.  The normalization of sound, $c_\mathrm{s}$ and tube, $c_\mathrm{T}$, speeds in the attenuating coefficients (\ref{eq:kappa}), contained in dispersion equation (\ref{eq:dispeq}), requires the values of both the reduced plasma betas $\tilde{\beta}_\mathrm{i,e} = c_\mathrm{i,e}^2/v_\mathrm{Ai,e}^2$ and the magnetic fields ratio $b = B_\mathrm{e}/B_\mathrm{i}$.  In the case of a twisted magnetic flux tube, for normalizing the local Alfv\'en speed (\ref{eq:omegaAi}), along with the fixed input parameters $\eta = \rho_\mathrm{e}/\rho_\mathrm{i}$ and $b_\mathrm{twist} = B_\mathrm{e}/B_{\mathrm{i}z}$, one needs to define the parameter $\varepsilon \equiv B_{\mathrm{i}\phi}(a)/B_{\mathrm{i}z}(a) = Aa/B_{\mathrm{i}z}$ that characterizes the twisted magnetic field inside the tube.  When the external magnetic field is twisted too, its normalization requires an additional $\varepsilon$, equal to $B_{\mathrm{e}\phi}(a)/B_{\mathrm{e}z}(a) = B_\phi/B_z$, which will have a subscript `2', while the similar parameter characterizing the twist of the internal magnetic field will possesses the subscript `1'.  After these preliminary notes, we can report the results of numerical computations, carried out for the two values of the density contrast, $\eta$, equal to $0.675$ and $0.403$, respectively.

\subsection{Kink mode propagation characteristics at a density contrast of $0.675$}
\label{subsec:eta=0675}
We begin with the simplest jet's model as an untwisted axially moving flux tube (the left column in Figure~\ref{fig:fig4}).
The ordering of basic speeds (sound, Alfv\'en, and tube one) in both media is as follows:
\[
    c_\mathrm{Ti} < c_\mathrm{Te} < v_\mathrm{Ai} < c_\mathrm{si} < c_\mathrm{se} < v_\mathrm{Ae}.
\]
According to \cite{Cally86} (see TABLE I there), at this ordering the kink ($m = 1$) mode, propagating in a static ($v_0 = 0$) tube, would be a pseudo surface/body wave of B$^{-}_{-}$ type.  In addition to $\eta = 0.675$, the other input parameters for the numerical task are: $\tilde{\beta}_\mathrm{i} = 1.2946$, $\tilde{\beta}_\mathrm{e} = 0.6836$, and $b = 1.184$.  At Alfv\'en Mach number $M_\mathrm{A} = 0$ (static flux tube) the anticipated normalized kink speed (\ref{eq:kinkspeed}) has a magnitude of $1.1977$.  With these input parameters the numerical computations of dispersion relation (\ref{eq:dispeq}) confirm that the kink mode traveling on the tube is a pseudo surface/body wave possessing, at $k_z a \ll 1$, a normalized phase velocity equal to the normalized kink speed within 4 places after the decimal point.  In moving flux tube, at relatively small $M_\mathrm{A}$s, the kink speed splits into a pair of phase velocities \citep{Zhelyazkov12}, whose dispersion curves initially go almost parallel to each
\begin{figure}[!ht]
   \centerline{\hspace*{0.015\textwidth}
               \includegraphics[width=0.515\textwidth,clip=]{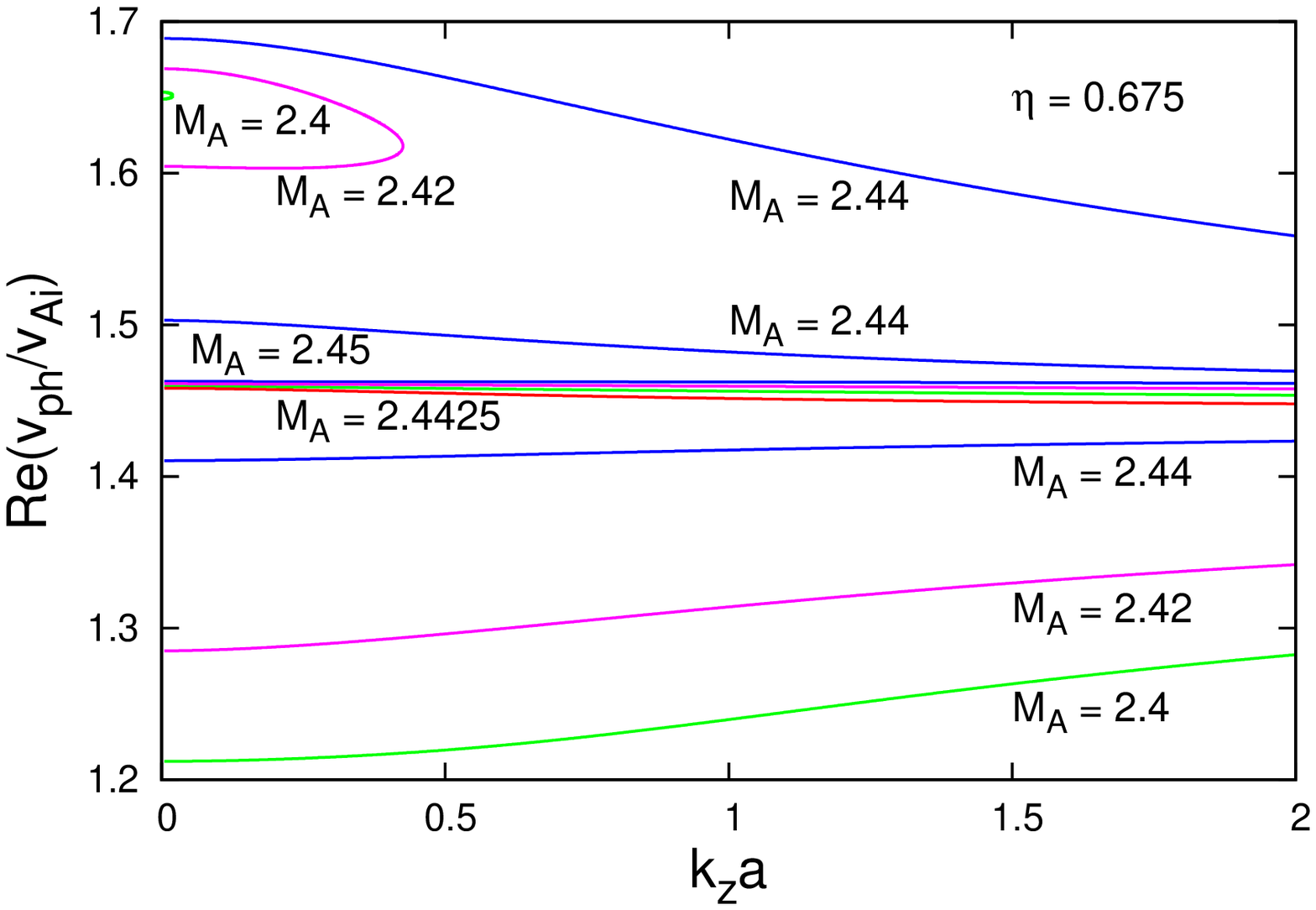}
               \hspace*{-0.03\textwidth}
               \includegraphics[width=0.515\textwidth,clip=]{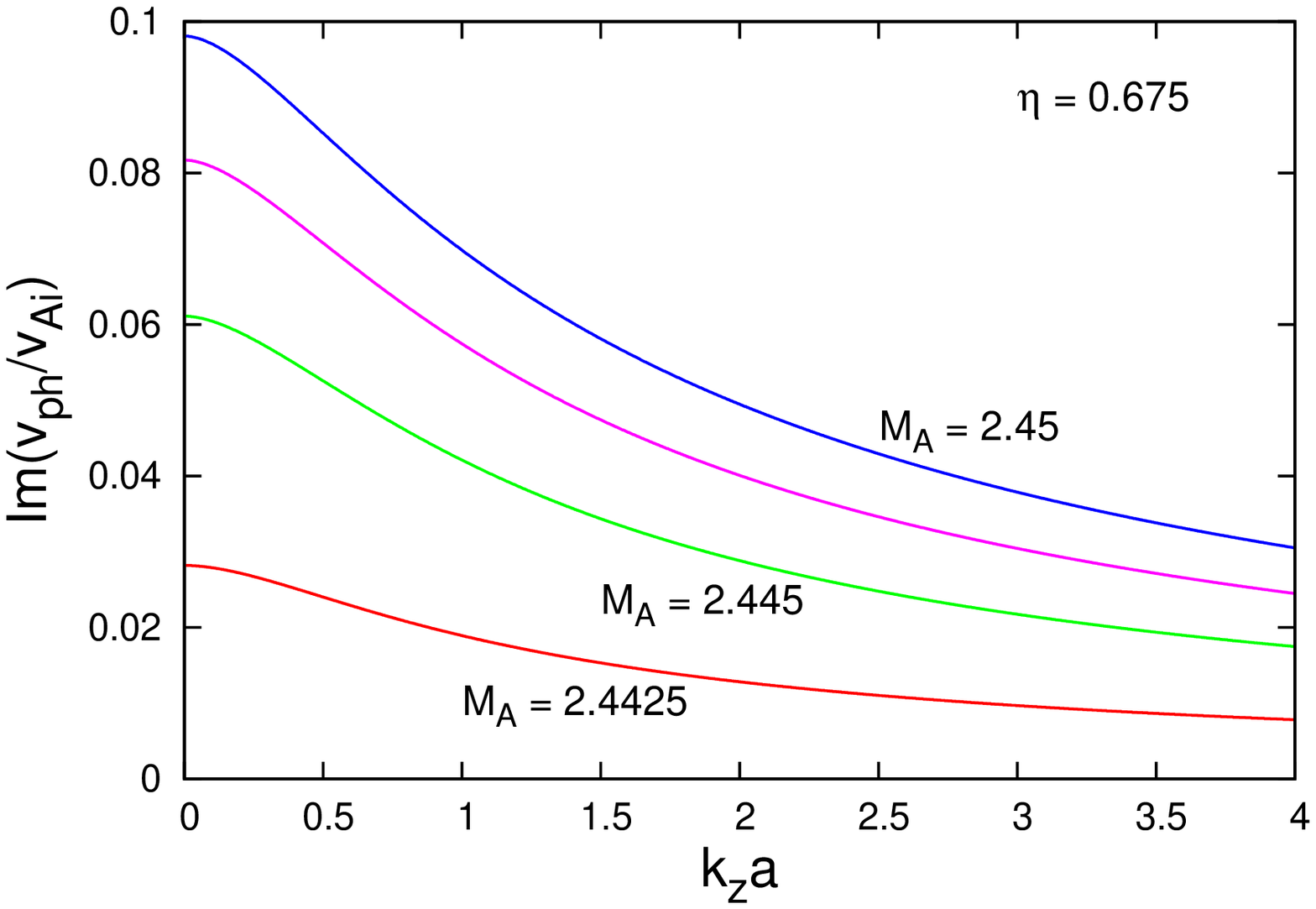}
              }
  \caption{(Left panel) Dispersion curves of stable and unstable kink ($m = 1$) mode propagating in a moving untwisted magnetic
  flux tube at $\eta = 0.675$, $b = 1.184$, and at various values of Alfv\'en Mach number $M_\mathrm{A}$.  The threshold Alfv\'en
  Mach number for KHI occurring is equal to $2.4425$ (red curve).  (Right panel) The normalized growth rates of the unstable kink
  mode for the same values of the input parameters.  The purple curve has been calculated at $M_\mathrm{A} = 2.4475$.}
   \label{fig:fig5}
\end{figure}
other, but for higher Alfv\'en Mach numbers, when one reaches the region of expected $M_\mathrm{A} > 2.44$ for occurring of KHI
according to the criterion (\ref{eq:criterion}), their behavior become completely different.  Look, for example, at the green and
purple curves labeled by $M_\mathrm{A} = 2.4$ and $M_\mathrm{A} = 2.42$ in the left panel of Figure~\ref{fig:fig5}.  The low-speed curves have more or less normal move while the high-speed ones turn over at some $k_z a$-values forming semi-closed dispersion curves.  The instability arises at the threshold Alfv\'en Mach number equal to $2.4425$---indeed rather close to the predicted value.  This value of $M_\mathrm{A}$ tells us that the KHI should emerge at a critical speed of the jet equal to $323.6$~km\,s$^{-1}$, which is less than the average jet speed of $332$~km\,s$^{-1}$ (see Table 1).  The marginal red curves divides the 1D $M_\mathrm{A}$-space into two regions: for all $M_\mathrm{A} < 2.4425$ the kink wave is a stable MHD mode, while in the opposite case it becomes unstable; alongside the marginal red curve in the left panel of Figure~\ref{fig:fig5}, one can see other three dispersion curves (in green, purple, and blue colors) presenting the unstable kink mode at the corresponding Alfv\'en Mach numbers.  The normalized growth rates of the same clutch of unstable kink waves are presented in the right panel of Figure~\ref{fig:fig5}.

For the computation of unstable kink ($m = 1$) mode propagating in axially moving twisted flux tube (the middle column in
Figure~\ref{fig:fig4}), our choice for the magnetic field twist parameter is $\varepsilon = 0.025$.  The other two input
\begin{figure}[!ht]
   \centerline{\hspace*{0.015\textwidth}
               \includegraphics[width=0.515\textwidth,clip=]{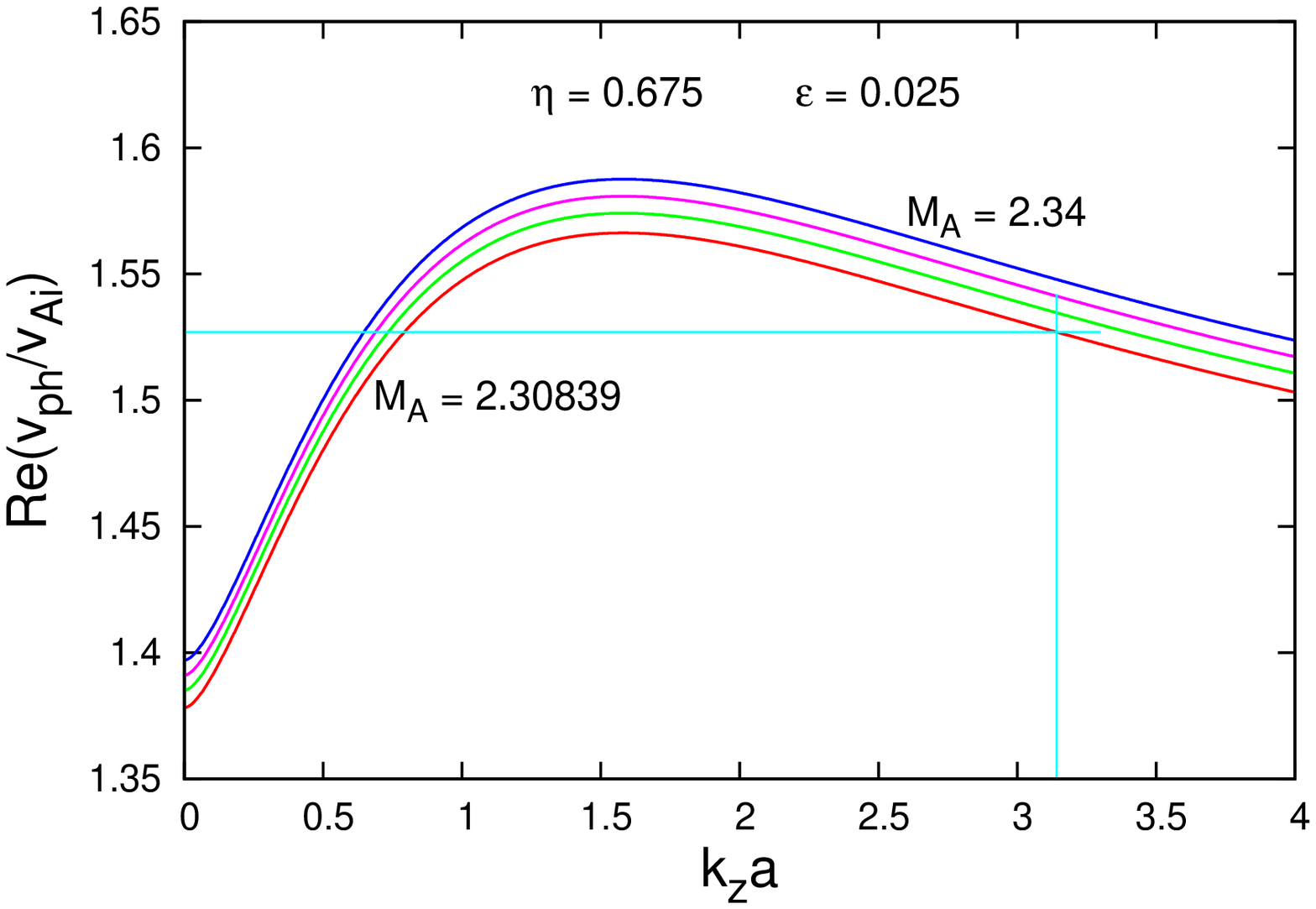}
               \hspace*{-0.03\textwidth}
               \includegraphics[width=0.515\textwidth,clip=]{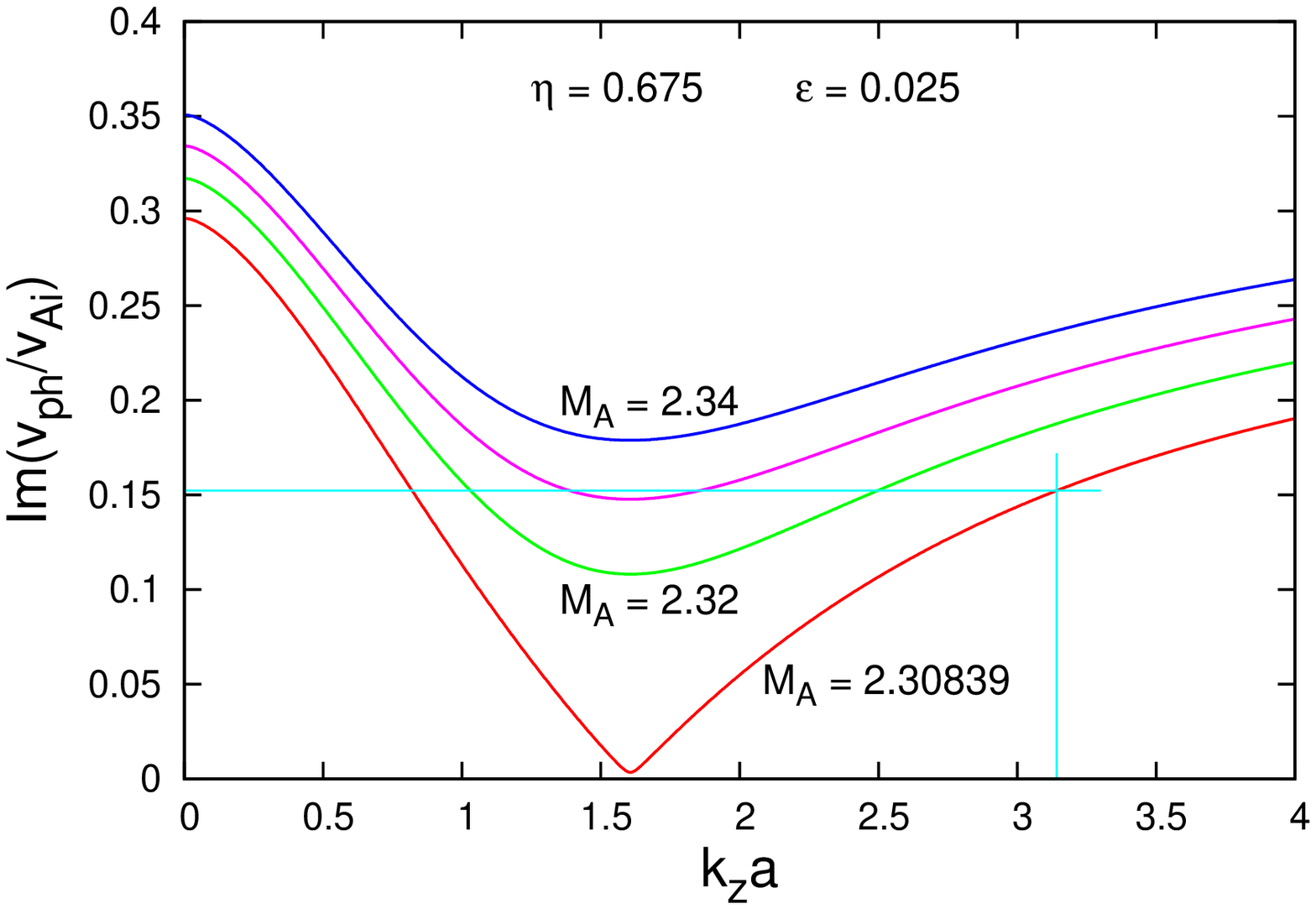}
              }
  \caption{(Left panel) Dispersion curves of unstable kink ($m = 1$) mode propagating in a moving twisted magnetic flux tube at
  $\eta = 0.675$, $b = 1$, and $\varepsilon = 0.025$ and at four various values of Alfv\'en Mach number $M_\mathrm{A}$.  The
  threshold Alfv\'en Mach number for KHI occurring is equal to $2.30839$ (red curve).  (Right panel) The normalized growth rates
  of the unstable kink mode for the same values of the input parameters.  The purple curve has been calculated at
  $M_\mathrm{A} = 2.33$.}
   \label{fig:fig6}
\end{figure}
parameters are $\eta = 0.675$ and $b = 1$.  The dispersion curves and corresponding normalized growth rates of the unstable mode for four values of the Alfv\'en Mach number are displayed in Figure~\ref{fig:fig6}.  There is an unexpected peculiarity---the threshold Alfv\'en Mach number for instability arising turns out to be lower than that in untwisted flux tube, notably equal to $2.30839$, which yields a critical flow speed of ${\approx}308$~km\,s$^{-1}$---$24$~km\,s$^{-1}$ less than the average jet speed.  An extensive study of KHI in an EUV jet situated on the west side of NOAA AR 10938 and observed on board \emph{Hinode\/} on 2007 January 15/16 showed just the opposite inequality: $2.4068$ vs $2.354327$, or $114.4$~km\,s$^{-1}$ vs $112$~km\,s$^{-1}$ \citep{Zhelyazkov16}.

It is curious to see what will be the wave growth rate, $\gamma_\mathrm{KH}$, instability developing time, $\tau_\mathrm{KH} = 2\pi/\gamma_\mathrm{KH}$, and the kink wave phase velocity, $v_\mathrm{ph}$, for a given wavelength.  Bearing in mind that our jet has a width of $4$~Mm and height of $152$~Mm, if we chose $\lambda_\mathrm{KH} = 4$~Mm to be a reasonable wavelength of the unstable mode, then the aforementioned instability parameters, determined by the crossed points of cyan lines and red marginally dispersion and growth rate curves at $k_z a = 3.141592$ in Figure~\ref{fig:fig6}, are:
\[
    \gamma_\mathrm{KH} = 31.7 \times 10^{-3}~\mathrm{s}^{-1}, \quad \tau_\mathrm{KH} = 198.4~\mathrm{s} = 3.3~\mathrm{min}, \quad
    \mbox{and} \quad v_\mathrm{ph} = 202.3~\mathrm{km}\,\mathrm{s}^{-1},
\]
\begin{figure}[!ht]
   \centerline{\hspace*{0.015\textwidth}
               \includegraphics[width=0.515\textwidth,clip=]{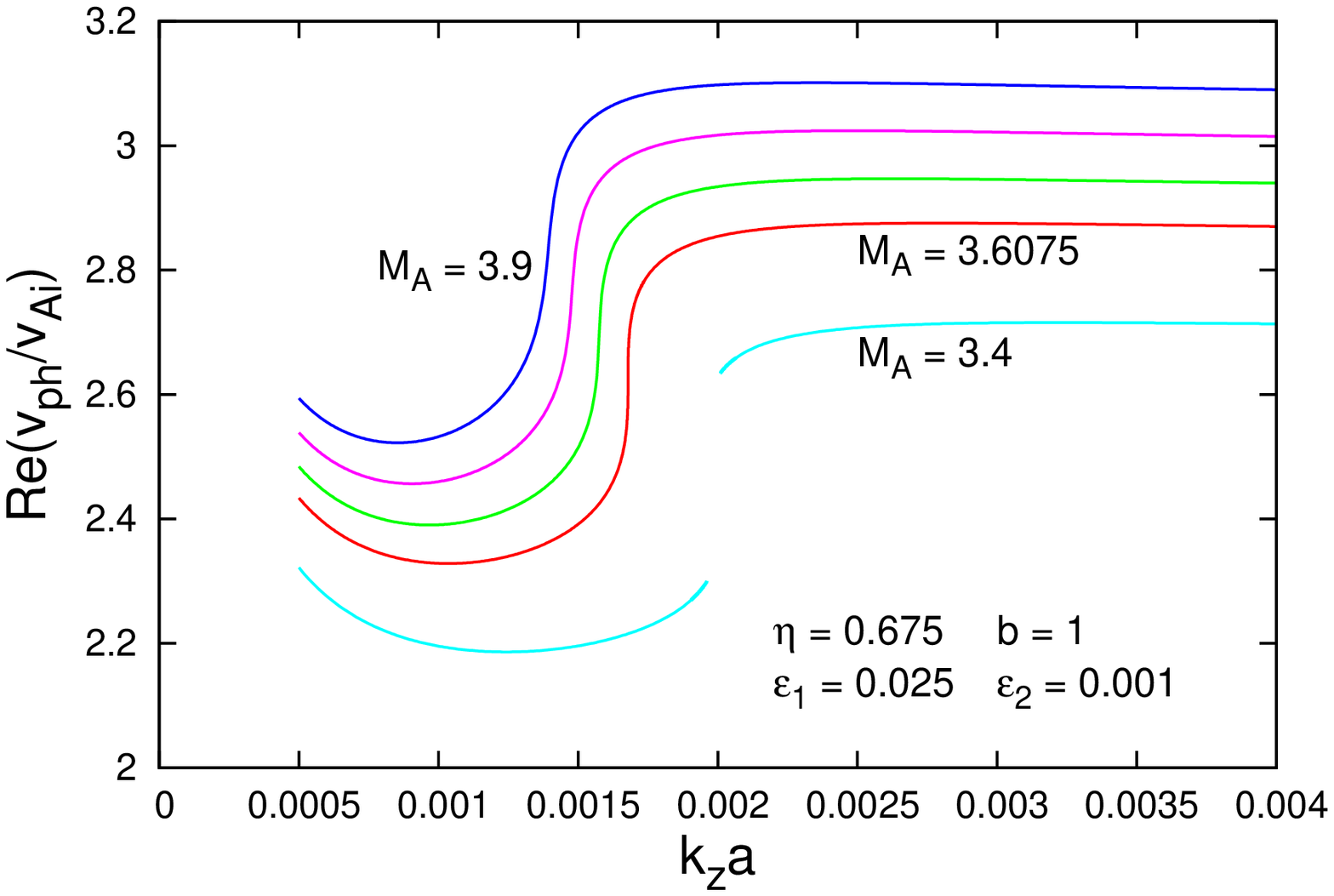}
               \hspace*{-0.03\textwidth}
               \includegraphics[width=0.515\textwidth,clip=]{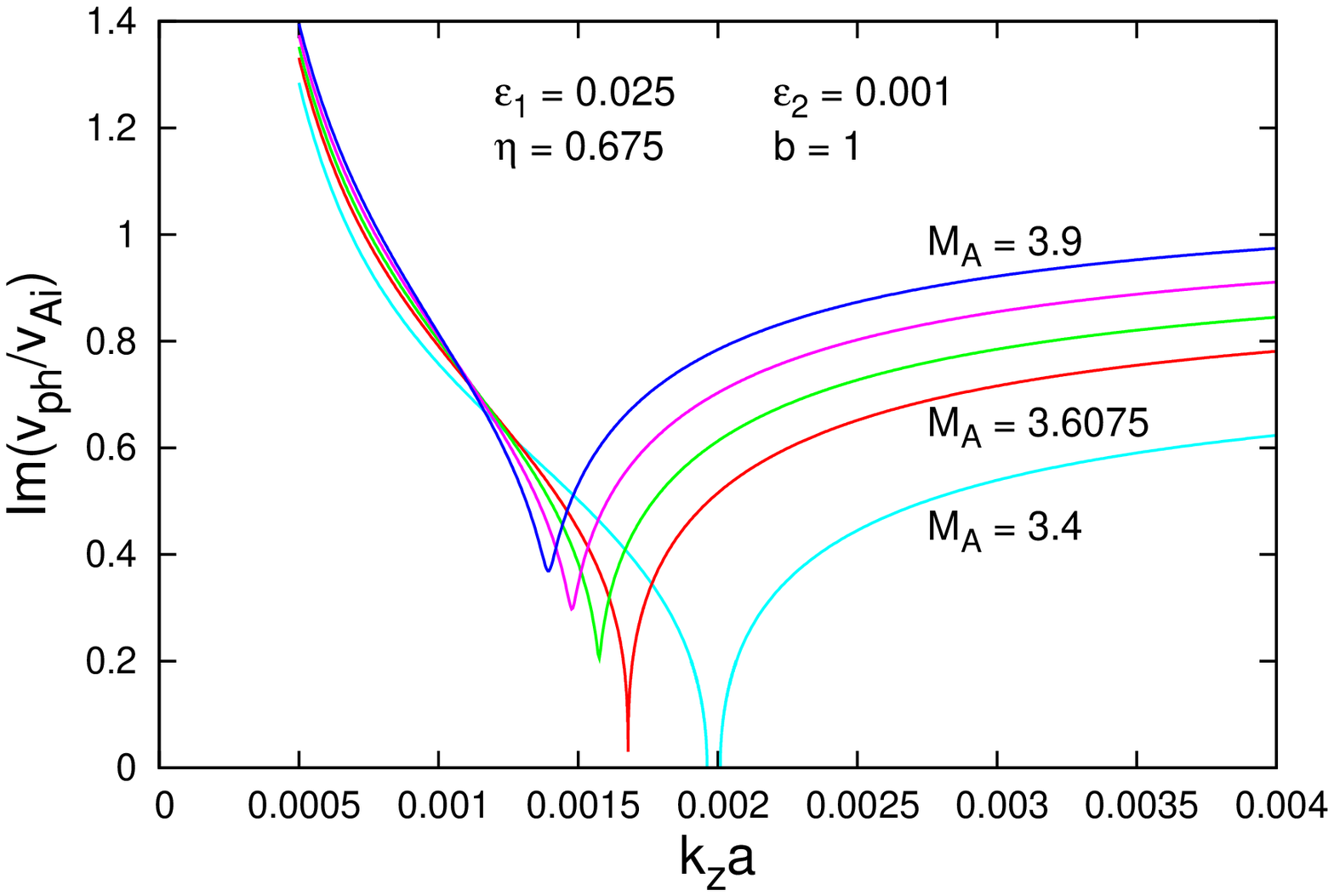}
              }
  \caption{(Left panel) Dispersion curves of unstable kink ($m = 1$) mode propagating in a moving twisted magnetic flux tube
  surrounded by a twisted magnetized plasma at
  $\eta = 0.675$, $b = 1$, $\varepsilon_1 = 0.025$, $\varepsilon_2 = 0.001$ and at five various values of Alfv\'en Mach number
  $M_\mathrm{A}$.  The two cyan curves labeled by $M_\mathrm{A} = 3.4$ are associated with the propagation of spurious unstable kink ($m = 1$)  mode.  The threshold Alfv\'en Mach number for KHI occurring is equal to $3.6075$ (red curve).  (Right panel) The normalized growth rates of the unstable and spurious unstable kink mode for the same values of the input parameters.  The green and purple curves have been calculated at $M_\mathrm{A} = 3.7$ and $3.8$, respectively.}
   \label{fig:fig7}
\end{figure}
respectively.  As seen, the KHI evolution time of $3.3$~min is much less than the jet lifetime of $16$~min (see Table 1 in
\citealp{Joshi17}).  Note that the two cross points in Figure~\ref{fig:fig6} can be considered as a ``phase portrait'' of KHI in the dimensionless phase velocity--wavenumber-plane.

In the most complicated case when the external magnetic field is also twisted (the right column in Figure~\ref{fig:fig4}), as we have mentioned, it is necessary to introduce two magnetic field twist parameters, $\varepsilon_1$ for the internal field, and
$\varepsilon_2$ for the external one, respectively.  We take $\varepsilon_1$, as in the previous case, to be equal to $0.025$; our choice for $\varepsilon_2$ is $0.001$.  Thus, with $\eta = 0.403$, $b = 1$, $\varepsilon_1 = 0.025$, and $\varepsilon_2 = 0.001$,
the solutions to the wave dispersion relation (\ref{eq:dbletwdispeq}) of the kink ($m = 1$) mode at five values of the Alfv\'en Mach number are graphically presented in Figure~\ref{fig:fig7}.  Here, we are faced with a distinctly different issue, namely at
relatively low Alfv\'en Mach numbers, in the very long wavelength limit, $k_z a \ll 1$, one appears two branches of the dispersion curve separated by a gap (see the cyan curves in Figure~\ref{fig:fig7}).  With increasing the magnitude of the Alfv\'en Mach number, that gap becomes narrower and at some $M_\mathrm{A}$ the two branches merge forming a continuous dispersion curve---this is the marginal dispersion curve and the corresponding Alfv\'en Mach number is the threshold one for appearance of the KHI---in our case its value is $3.6075$.  We would like to notice that the red normalized growth rate curve has no cusp at $k_z a = 0.00168$---it is a normal smooth curve.  For larger values of $k_z a$ both the dispersion and growth rate curves go with gradually increasing magnitudes.  But at such high threshold Alfv\'en Mach number $M_\mathrm{A} = 3.6075$ and Alfv\'en speed of $132.46$~km\,s$^{-1}$, the required flow speed for instability onset is equal to $478$~km\,s$^{-1}$---a value which is inaccessible by the jet under consideration.  Hence, even at a very small twisted external magnetic field suppresses the KHI occurrence.

\subsection{Kink mode propagation characteristics at a density contrast of $0.403$}
\label{subsec:eta=0403}
With the lower value of the density contrast, $\eta = 0.403$, the magnitudes of sound speeds in both media are unchanged, but those of the Alfv\'en speeds are changed and now the ordering of four basic speeds is
\[
    c_\mathrm{Ti} < v_\mathrm{Ai} < c_\mathrm{Te} < c_\mathrm{si} < c_\mathrm{se} < v_\mathrm{Ae},
\]
which implies (see TABLE I in \citealp{Cally86}) that the kink mode propagating in a rest untwisted flux tube should be a surface
wave of S$^{-}_{+}$ type.  The normalized magnitude of the kink speed is $c_\mathrm{k}/v_\mathrm{Ai} = 2.03457$.  The mode will become unstable against KHI if $M_\mathrm{A} > 4.496$.  The input parameters for the numerical solution of dispersion equation (\ref{eq:dispeq}) now are: $\eta = 0.403$, $\tilde{\beta}_\mathrm{i} = 5.6899$, $\tilde{\beta}_\mathrm{e} = 0.6282$, and $b = 2.192$.  With these input values, at $M_\mathrm{A} = 0$ (static magnetic flux tube) numerical computations confirm that the kink ($m = 1$) mode is a pure surface wave and in the long wavelength limit, $k_z a \ll 1$, recover the normalized kink speed of $2.03457$ within three places behind the decimal point.  In searching that Alfv\'en Mach number at which the KHI will merge, we use another strategy, notably starting the computations with a higher than the predicted $4.496$ value of $M_\mathrm{A}$, and latter on decreasing it (with small steps) until reach the marginal dispersion and normalized growth rate curves---the results are
\begin{figure}[!ht]
   \centerline{\hspace*{0.015\textwidth}
               \includegraphics[width=0.515\textwidth,clip=]{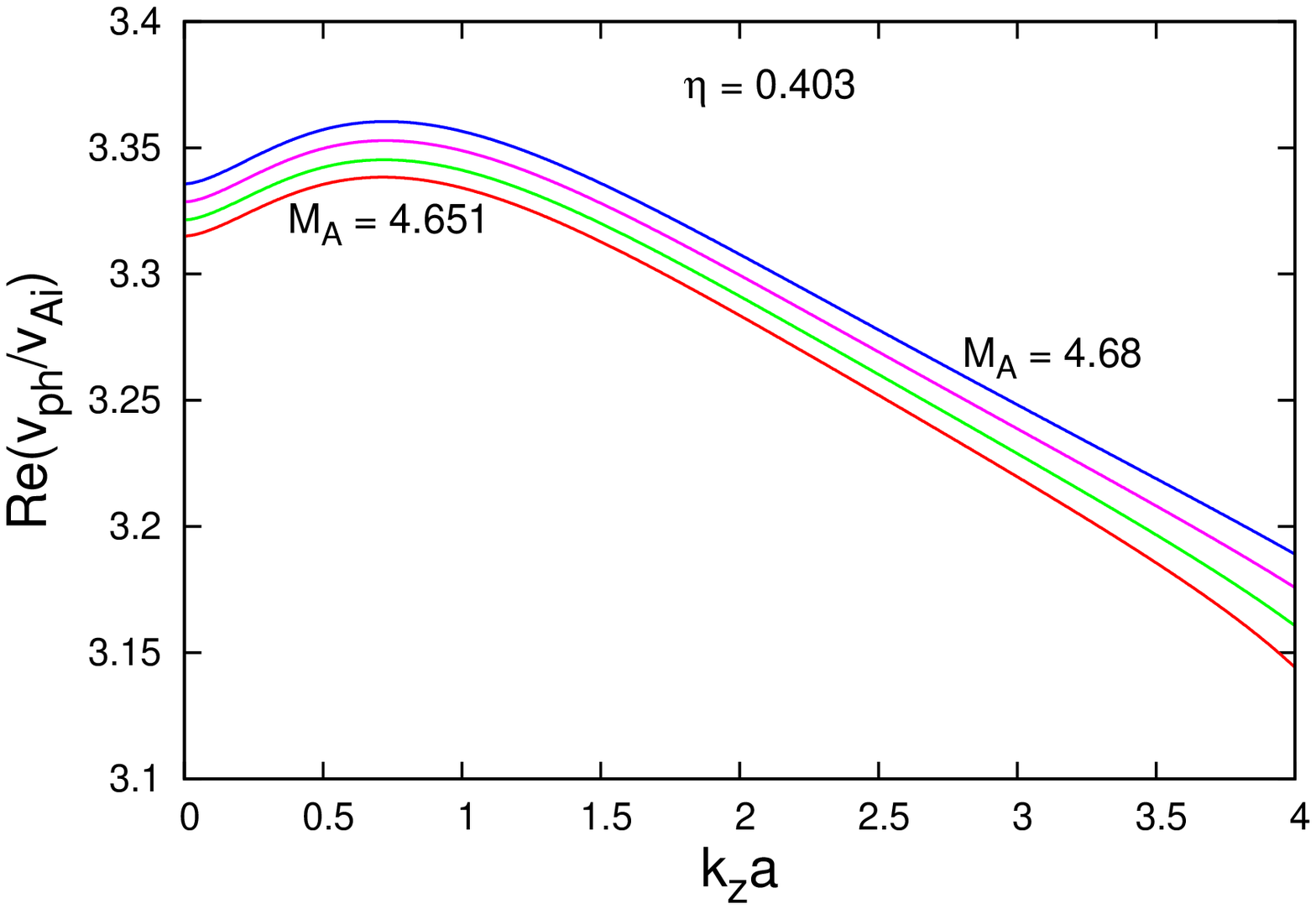}
               \hspace*{-0.03\textwidth}
               \includegraphics[width=0.515\textwidth,clip=]{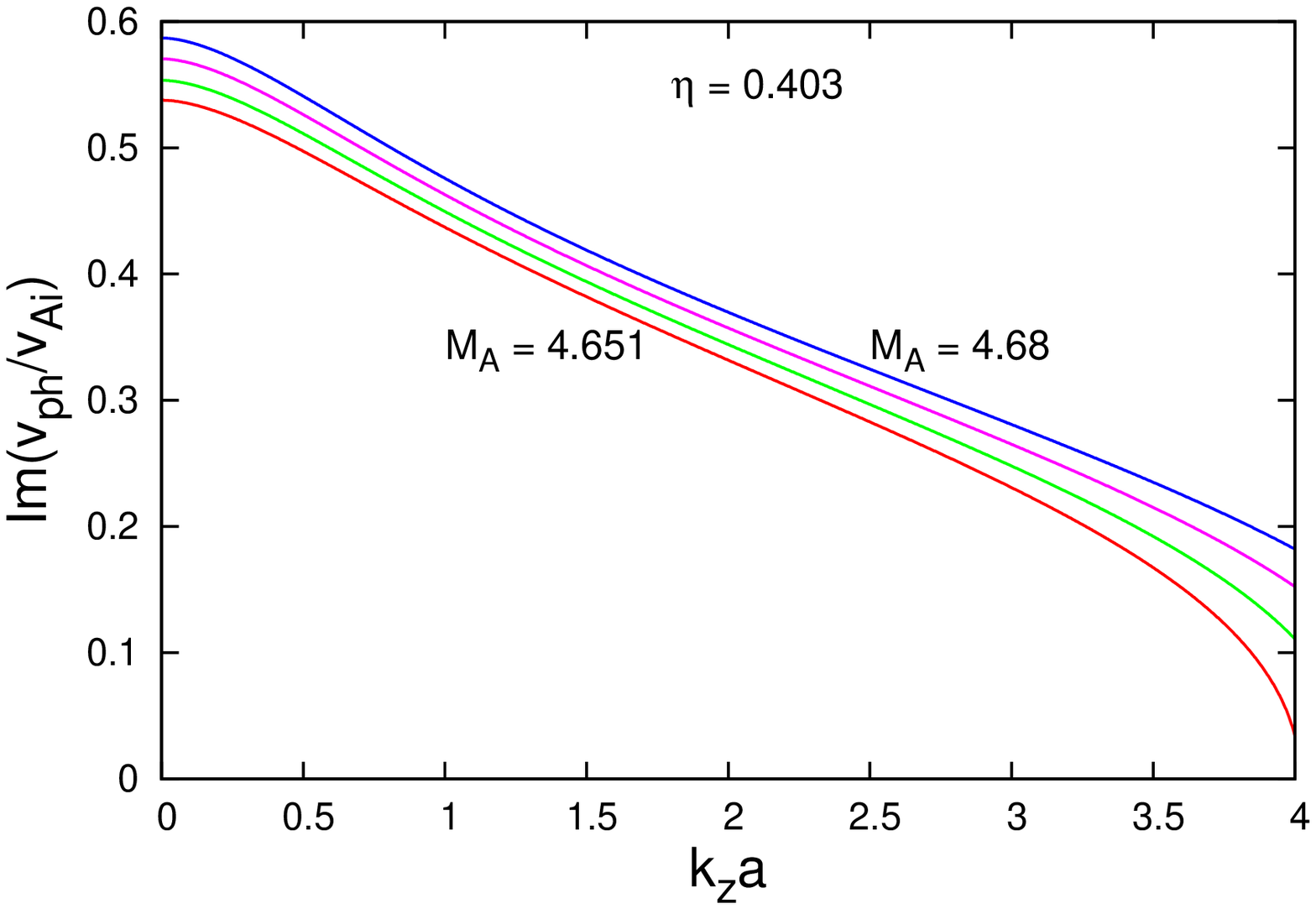}
              }
  \caption{(Left panel) Dispersion curves of unstable kink ($m = 1$) mode propagating in a moving untwisted magnetic
  flux tube at $\eta = 0.403$, $b = 2.19$, and at various values of $M_\mathrm{A}$.  The threshold Alfv\'en
  Mach number for KHI occurring is equal to $4.651$ (red curve).  (Right panel) The normalized growth rates of the unstable kink
  mode for the same values of the input parameters.  The green and purple curve have been calculated at $M_\mathrm{A} = 4.66$
  and $4.67$, respectively.}
   \label{fig:fig8}
\end{figure}
\begin{figure}[!ht]
   \centerline{\hspace*{0.015\textwidth}
               \includegraphics[width=0.515\textwidth,clip=]{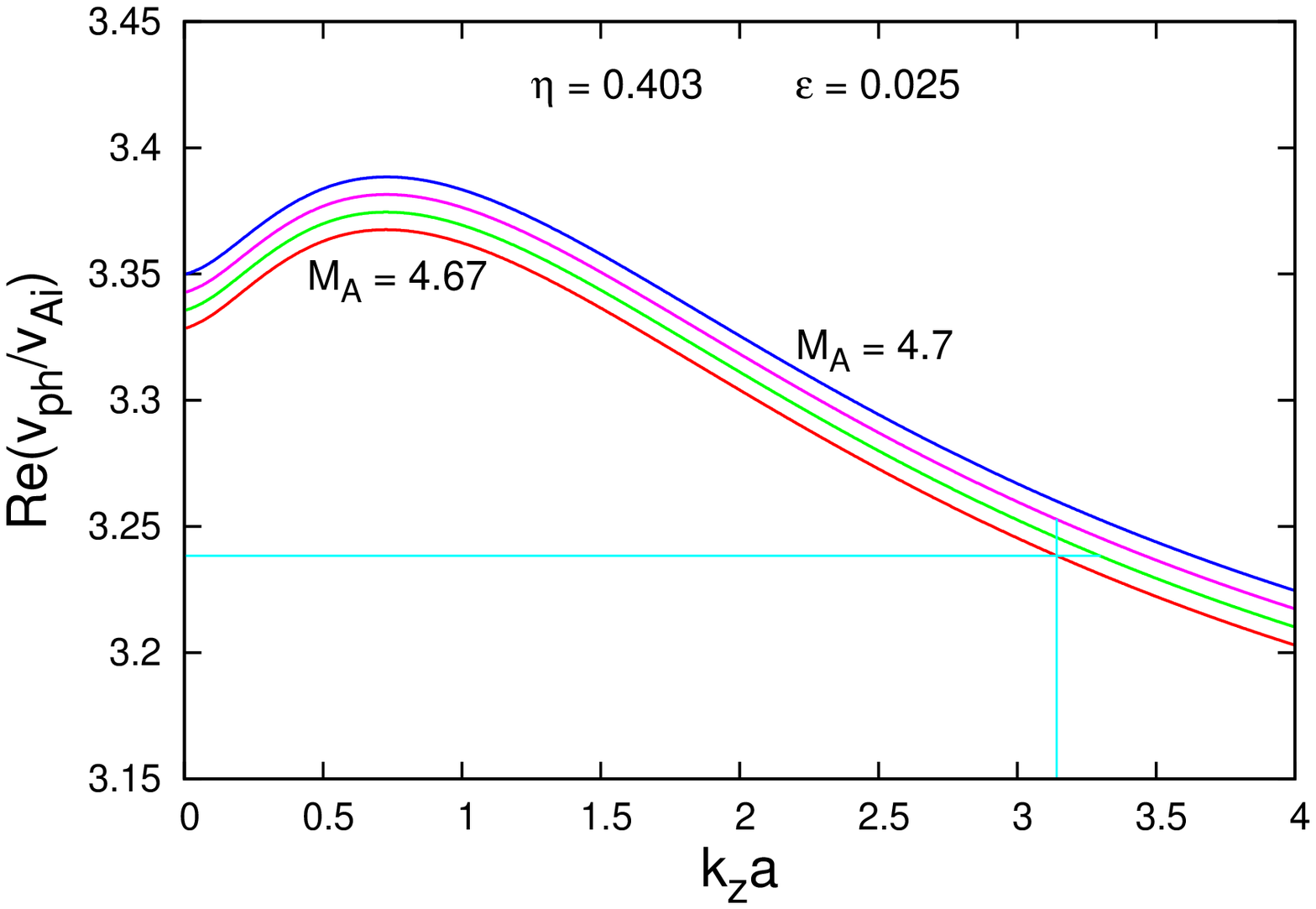}
               \hspace*{-0.03\textwidth}
               \includegraphics[width=0.515\textwidth,clip=]{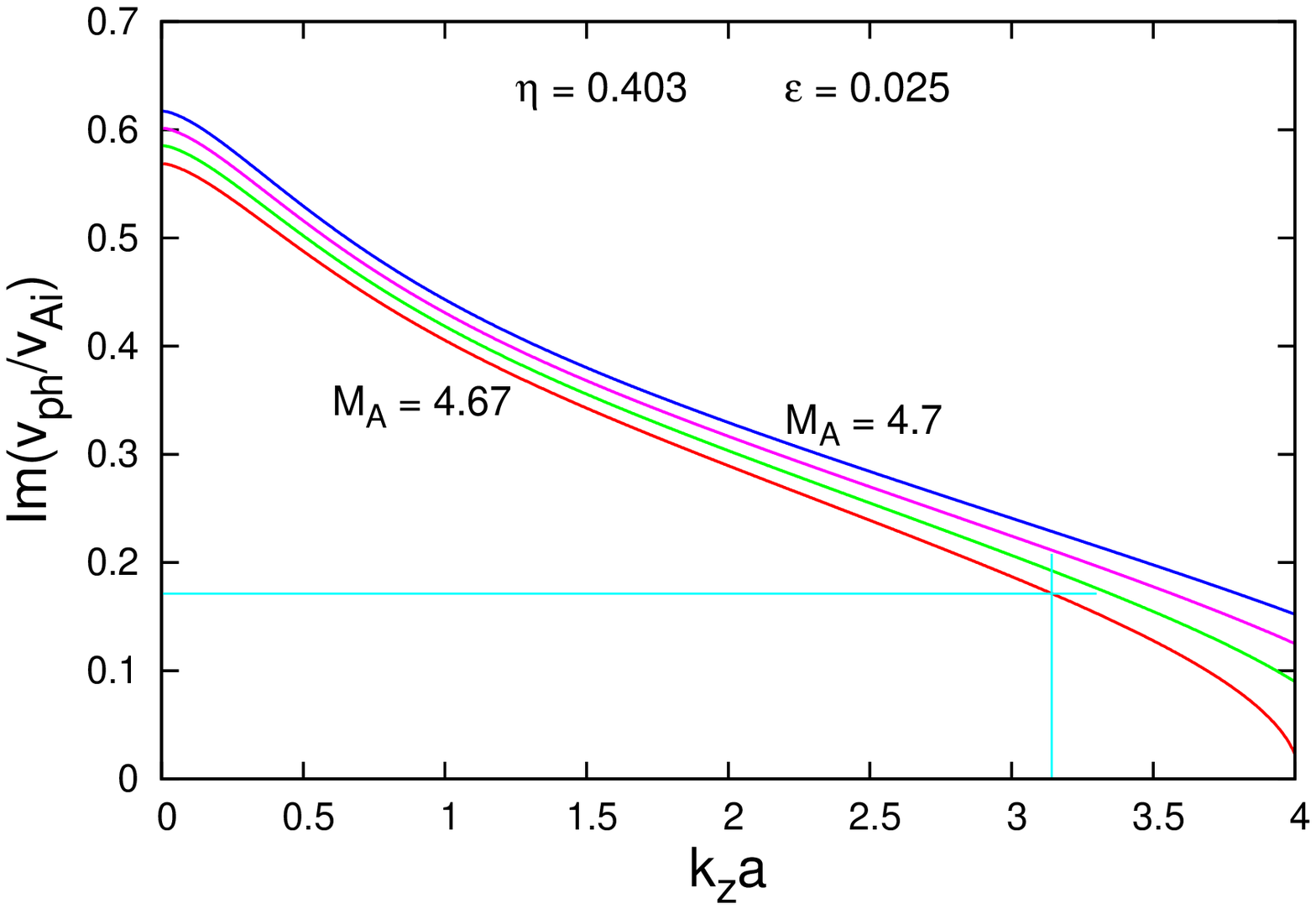}
              }
  \caption{(Left panel) The same as in Figure~\ref{fig:fig6}, but for $\eta = 0.403$ (cool environment) and $b = 2.19$.  The threshold $M_\mathrm{A}$ for KHI occurring is equal to $4.67$ (red curve).  (Right panel)  The green and purple curves have been calculated at $M_\mathrm{A} = 4.68$ and $4.69$, respectively.}
   \label{fig:fig9}
\end{figure}
shown in Figure~\ref{fig:fig8}.  The threshold Alfv\'en Mach number is equal to $4.651$ which means that with $v_\mathrm{Ai} = 63.2$~km\,s$^{-1}$ the critical flow velocity for KHI onset is $294$~km\,s$^{-1}$, which is lower than the average jet speed of $332$~km\,s$^{-1}$.

When the internal magnetic field is twisted and the external medium is cool magnetized plasma, we have to numerically solve dispersion Equation~(\ref{eq:twdispeq}), but calculating the external wave attenuation coefficient $\kappa_\mathrm{e}$ by using expression~(\ref{eq:kappa_e}).  With the input parameters $\eta = 0.403$, $b = 2.19$, and $\varepsilon = 0.025$, the numerical solutions for the unstable ($m = 1$) kink mode are displayed in Figure~\ref{fig:fig9}.  The threshold Alfv\'en Mach number for instability onset being equal to $4.67$ is higher than that for untwisted moving flux tube, but critical flow velocity (calculated with $v_\mathrm{Ai} = 63.18$~km\,s$^{-1}$) of $295$~km\,s$^{-1}$ is still lower that the average jet speed.  This means that the KHI should arise and at an instability wavelength $\lambda_\mathrm{KH} = 4$~Mm (that is, at $k_z a = 3.141592$) the instability
\begin{figure}[!ht]
   \centerline{\hspace*{0.015\textwidth}
               \includegraphics[width=0.515\textwidth,clip=]{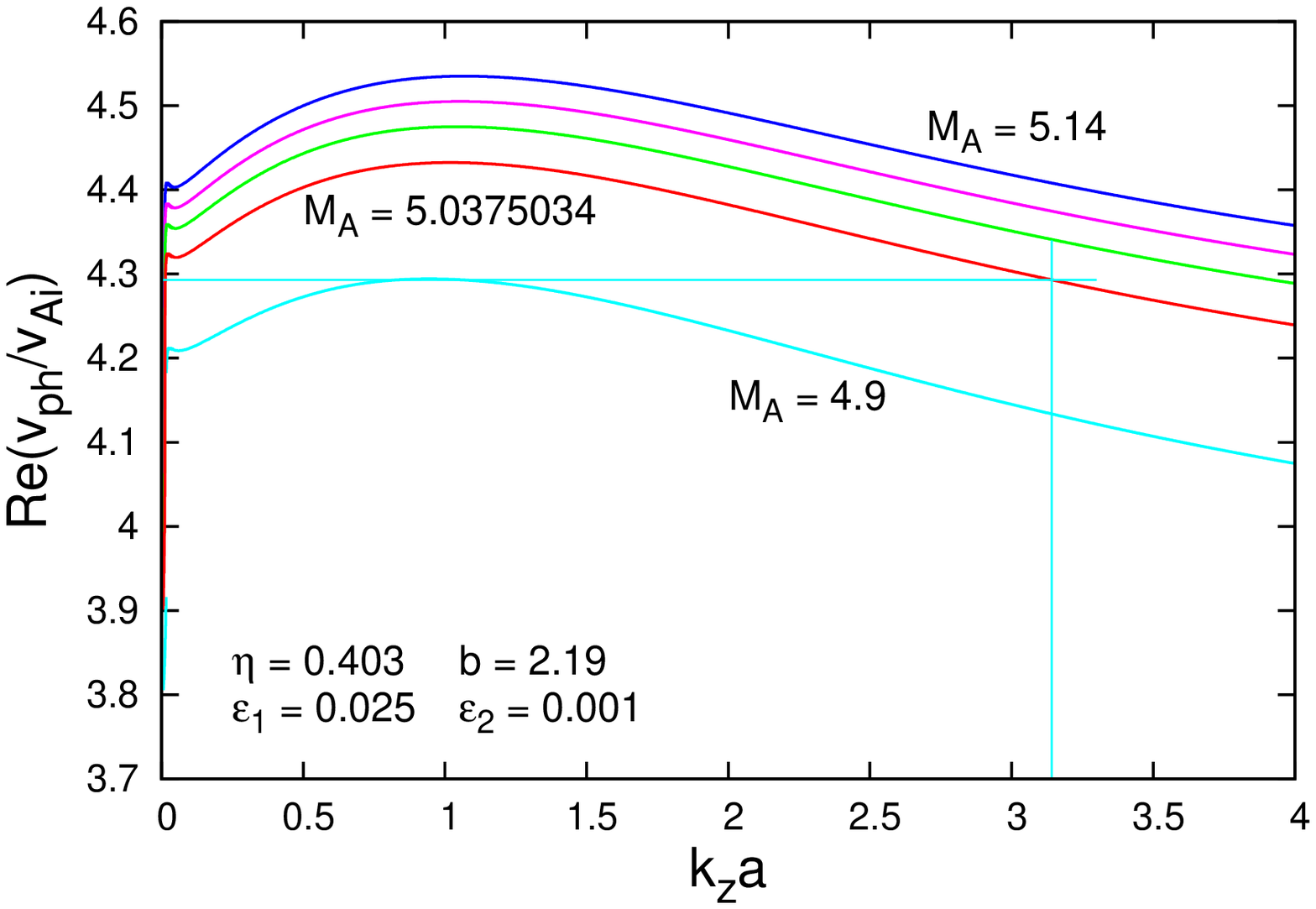}
               \hspace*{-0.03\textwidth}
               \includegraphics[width=0.515\textwidth,clip=]{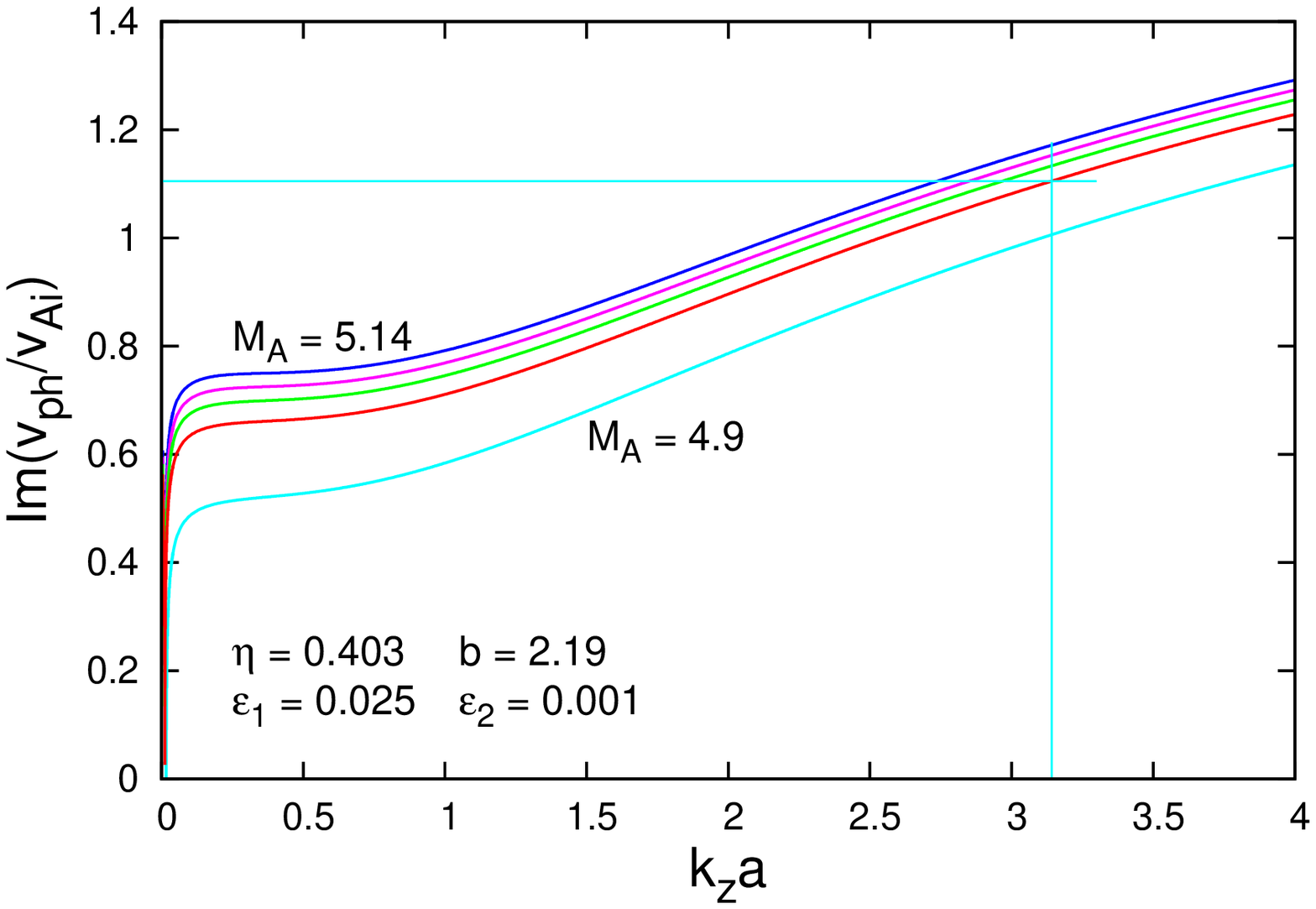}
              }
  \caption{(Left panel) Dispersion curves of unstable kink ($m = 1$) mode propagating in a moving twisted magnetic flux tube
  surrounded by a cool twisted magnetized plasma at $\eta = 0.675$, $b = 2.19$, $\varepsilon_1 = 0.025$, $\varepsilon_2 = 0.001$ and at five various values of $M_\mathrm{A}$.  The cyan curve labeled by $M_\mathrm{A} = 4.9$ simulates spurious unstable kink
  ($m = 1$) mode.  The threshold $M_\mathrm{A}$ for KHI occurring is equal to $5.0375034$ (red curve).  (Right panel) The
  normalized growth rates of the unstable and spurious unstable kink mode for the same values of the input parameters.  The green and purple curves have been calculated at $M_\mathrm{A} = 5.08$ and $5.11$, respectively.}
   \label{fig:fig10}
\end{figure}
wave growth rate, instability developing/evolution time, and mode phase velocity are:
\[
    \gamma_\mathrm{KH} = 17.0 \times 10^{-3}~\mathrm{s}^{-1}, \quad \tau_\mathrm{KH} = 369.5~\mathrm{s} = 6.2~\mathrm{min}, \quad
    \mbox{and} \quad v_\mathrm{ph} = 204.6~\mathrm{km}\,\mathrm{s}^{-1},
\]
respectively.  It is immediately seen that now the KHI instability evolution time is $1.88$ times longer than that in the case of $\eta = 0.675$, while the wave phase velocities are of the same order.  In this case, when the environment is a cool medium, looking at the right panel in Figure~\ref{fig:fig9}, one recognizes that there exists a peculiarity, notably the instability ceases at some dimensionless wavenumber---in our case is it equal to $4.017$.  If one needs a little bit shorter instability wavelength $\lambda_\mathrm{KH}$, that is bigger $k_z a$, one has to increase the magnitude of the threshold Alfv\'en Mach number.  This circumstance is not too dangerous because there is room for KHI onset at threshold $M_\mathrm{A}$ up to $5.25$.
	
\begin{figure}[!ht]
   \centerline{\hspace*{0.015\textwidth}
               \includegraphics[width=0.515\textwidth,clip=]{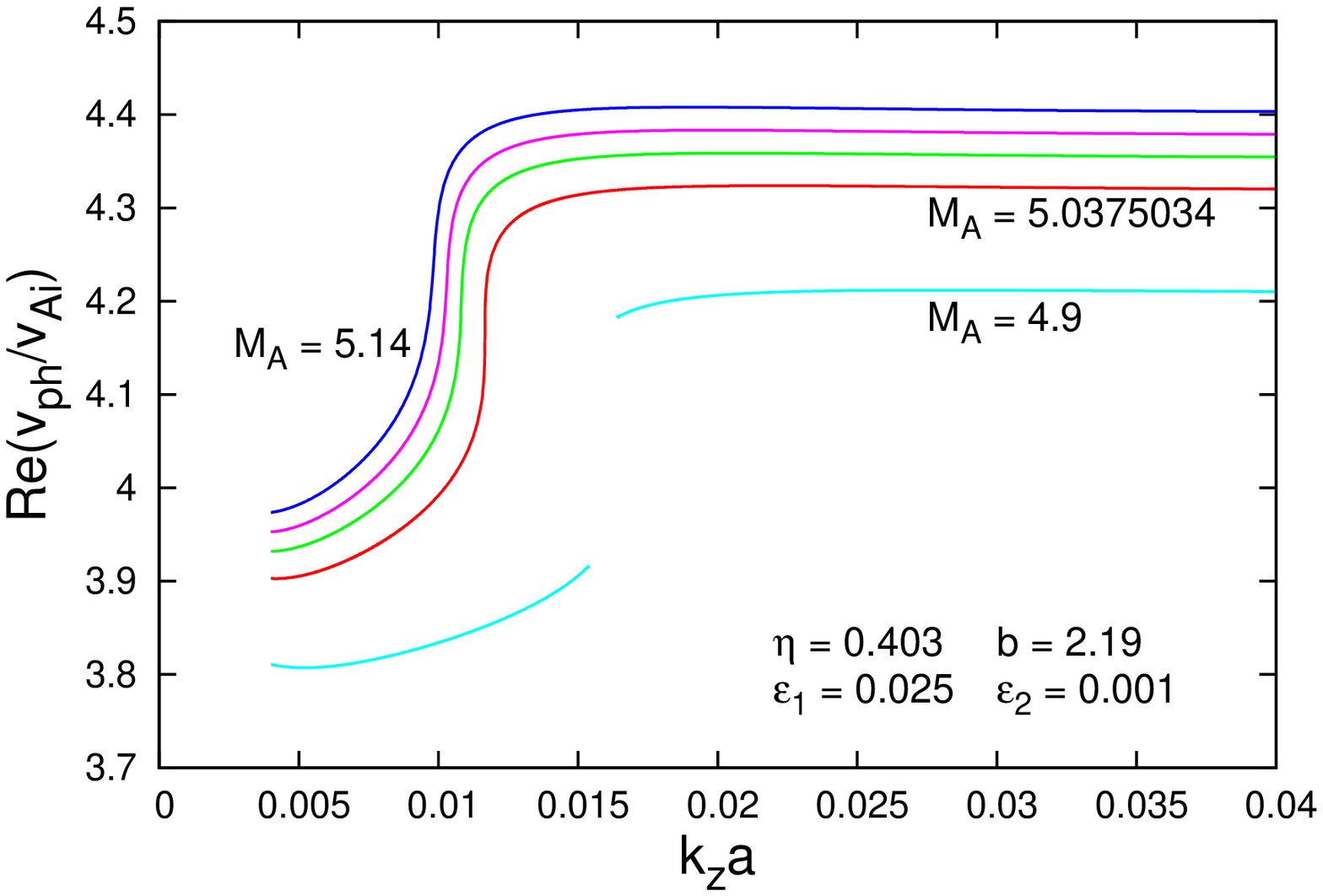}
               \hspace*{-0.03\textwidth}
               \includegraphics[width=0.515\textwidth,clip=]{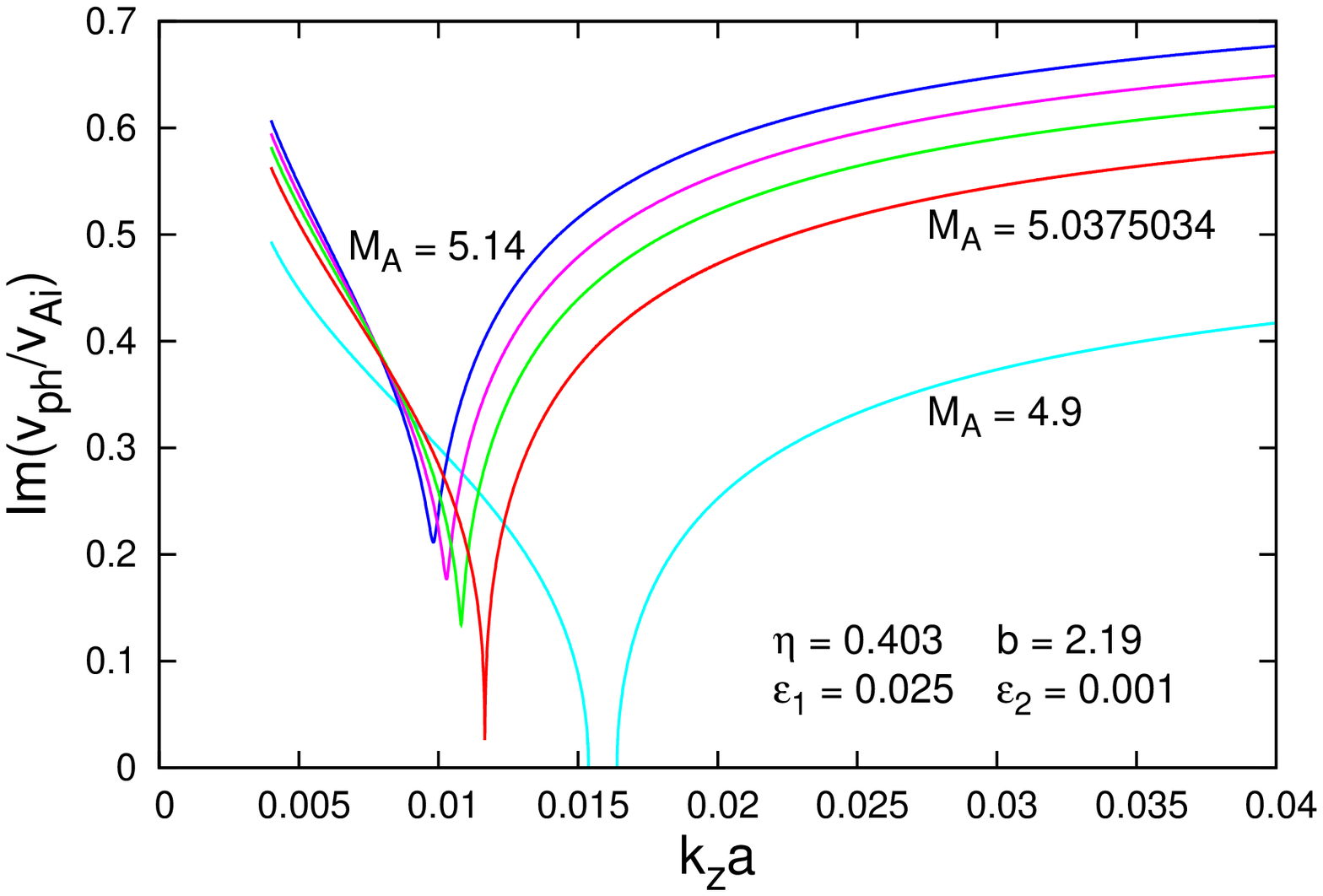}
              }
  \caption{(Left panel) A zoom of dispersion and growth rate curves of the kink ($m = 1$) mode as displayed in Figure~\ref{fig:fig10} at the very beginning of the horizontal $k_z a$-axis.}
   \label{fig:fig11}
\end{figure}

Twisted external magnetic field, as in the previous case of $\eta = 0.675$ when the environment was considered as incompressible medium, dramatically change dispersion characteristics of the kink mode.  Solving dispersion relation (\ref{eq:dbletwdispeq}) with $\eta = 0.403$, $b = 2.19$, $\varepsilon_1 = 0.025$, and $\varepsilon_2 = 0.001$, we obtain families of dispersion and normalized growth rate curves pictured in Figure~\ref{fig:fig10}.  The distinctly peculiar shape of the dispersion and growth rate curves at the very beginning of the $k_z a$-axis is seen in Figure~\ref{fig:fig11}.  That figure is similar to Figure~\ref{fig:fig7}, but
note that the cusp-like growth rate curves are shifted on one order to the right.  Cyan-colored curves in both figures (\ref{fig:fig10} and \ref{fig:fig11}) calculated at $M_\mathrm{A} = 4.9$ are associated with a spurious unstable kink mode---a
really unstable kink wave arises at $M_\mathrm{A} = 5.0375034$ (red curves in the two figures).  By contrast to the previous case of $\eta = 0.675$, now the KHI instability of the kink mode can merge at a flow velocity of ${\approx}318$~km\,s$^{-1}$ which is lower that the average jet speed of $332$~km\,s$^{-1}$.  Going back to Figure~\ref{fig:fig10}, it is intriguing to see how the twist of the external magnetic filed changes the KHI characteristics of the kink mode at the instability wavelength $\lambda_\mathrm{KH} = 4$~Mm compared with those calculated for the previous case, when the internal magnetic field is twisted only.  The cross points of vertical cyan lines, plotted at $k_z a = 3.141592$, with corresponding marginal dispersion and normalized growth rate curves yield the following values:
\[
    \gamma_\mathrm{KH} = 109.7 \times 10^{-3}~\mathrm{s}^{-1}, \quad \tau_\mathrm{KH} = 57.3~\mathrm{s} \cong 1~\mathrm{min}, \quad
    \mbox{and} \quad v_\mathrm{ph} = 271~\mathrm{km}\,\mathrm{s}^{-1}.
\]
We see that now the instability evolution time, $\tau_\mathrm{KH}$, is much shorter (${\cong}1$~min), that is KHI is faster, and the phase velocity of the unstable kink ($m = 1$) mode is higher, $271$ vs $205$~km\,s$^{-1}$.

\section{Conclusion}
\label{sec:conclusion}

We have explored the possibilities for arising the KHI of the kink ($m = 1$) mode in a fast EUV jet observed by \emph{SDO}/AIA in the NOAA active region 12035 on April 16, 2014.  We have modeled the jet as a moving cylindrical magnetic flux tube of radius $a$ at three different magnetic fields' topologies, namely untwisted homogeneous magnetic fields inside and outside the tube, twisted internal magnetic field and untwisted one in the environment, and both twisted magnetic fields.

Depending on the value of the density contrast (be it $\eta = 0.675$ or $0.403$), the magnitude of the background magnetic field $B_\mathrm{e}$, and magnetic field topology, we treat the internal and external media as compressible plasmas (when the moving tube is untwisted), or as incompressible internal and external media (at $\eta = 0.675$), or as incompressible internal and cool environment media (at $\eta = 0.403$) when one or both magnetic fields are twisted.

Studying the kink ($m = 1$) MHD mode propagation in an untwisted moving magnetic flux tube allows us to determine depending on the ordering of sound, Alfv\'en, and tube velocities in both media the nature of the propagating mode (pure surface or pseudo surface wave \citep{Cally86}) along with the value of the characteristic kink speed $c_\mathrm{k}$ (\ref{eq:kinkspeed})---both issues in a rest flux tube.  The normalized with respect $v_\mathrm{Ai}$ kink speed must be reproduced in the $k_z a \ll 1$ limit during the numerical solution of dispersion equation (\ref{eq:dispeq}) if the numerical code is correct.  The inclusion of jet speed via the Alfv\'en Mach number does not change the wave nature until the kink mode is stable, but splits the kink speed dispersion curves into a pair of curves whose behavior might be generally rather complex as illustrated in the left panel of Figure~\ref{fig:fig5}. The threshold $M_\mathrm{A}$ can be predicted on using criterion~(\ref{eq:criterion}), but its actual value is given by the computed unstable normalized wave phase velocity and growth rate curves (termed marginal ones and plotted in red color).  Numerically derived threshold Alfv\'en Mach numbers turn out to be very close to the predicted values for the two sets of input  data of both density contrasts.  If the critical jet velocity for KHI onset is accessible, one can, in principle, evaluate at a fixed wavelength the instability developing time and the corresponding wave phase velocity.  We note that when the kink mode becomes unstable both attenuation coefficients determining its spatial structure in the two media are complex quantities and it is logical such a wave to be called \emph{generalized surface mode}, that is, being neither pure surface, pseudo surface/body, or leaky wave.

Observations show that in the most cases solar atmosphere jets are more or less twisted.  That is why the focus of our study was to see how the magnetic field twist changes the picture. At $\eta = 0.675$ with $b = 1$ and $\varepsilon = 0.025$, the threshold $M_\mathrm{A}$, unexpectedly was found to be lower than the one for untwisted magnetic flux tube, namely $2.30839$ vs $2.4425$.  At the other $\eta = 0.403$, with $b = 2.19$ and $\varepsilon = 0.025$, we have just the opposite relation: KHI instability onset of the kink ($m = 1$) mode requires a little bit higher critical jet speed than that for untwisted tube: $295$ vs $294$~km\,s$^{-1}$.  It is instructive to note that the difference between the corresponding threshold Alfv\'en Mach numbers is more emphatic ($4.67$ vs $4.651$), but the slightly different Alfv\'en speeds ($63.18$ and $63.2$~km\,s$^{-1}$, respectively) make the critical flow velocities difference rather small---$1$~km\,s$^{-1}$ only.  Another important observation is the different shapes of normalized growth rate curves, more specifically, in the case when the environment is treated as a cool medium ($\eta = 0.403$), the growth rate curve is limited on the right-hand side of the horizontal axis in the Im($v_\mathrm{ph}/v_\mathrm{Ai})$--$k_z a$-plane.  In such a case, if one needs a wider $k_z a$ instability range, one must increase the threshold $M_\mathrm{A}$.  It is worth mentioning that irrespective of slightly different background magnetic fields ($7$ and $8$~G, respectively), the KHI developing times of the kink mode at $\lambda_\mathrm{KH} = 4$~Mm are not too distinct though at $\eta = 0.304$ the evolution time of $6.2$~min is approximately two times longer than the one at $\eta = 0.675$ being equal to $3.3$~min.  On the other side, the wave phase velocities of the unstable kink mode are very close: $202.3$~km\,s$^{-1}$ for $\eta = 0.675$ and $204.6$~km\,s$^{-1}$ at $\eta = 0.403$.

A definitely new achievement in our paper is the study of how the external twisted magnetic field affects the development of KHI. In that case, in addition to the internal magnetic field twist parameter, now subscribed by `1', that is, $\varepsilon_1 = 0.025$, we had to introduce a second twist parameter characterizing the external magnetic field, whose value is relatively small: $\varepsilon_2 = 0.001$.  The influence of the twisted external magnetic field $\vec{B}_\mathrm{e}$ on the instability onset critically depends upon the nature of surrounding coronal plasma.  If the environment is treated as incompressible plasma, that is, $\eta = 0.675$, the external twisted magnetic field suppresses the instability arising: with a threshold Alfv\'en Mach number equal to $3.6075$ (see Figure~\ref{fig:fig7}) and Alfv\'en speed $v_\mathrm{Ai} = 232.46$~km\,s$^{-1}$, the required critical flow velocity of ${\cong}478$~km\,s$^{-1}$ for the instability triggering is far beyond the average jet speed of $332$~km\,s$^{-1}$.  In the case when the surrounding medium is considered as cool plasma ($\eta = 0.403$), it turns out that the kink ($m = 1$) mode can become unstable at an accessible critical jet speed of ${\approx}318$~km\,s$^{-1}$.  Comparing the KHI evolution times of the kink mode propagating in single-twisted and double-twisted moving magnetic flux tube we establish that in the second case $\tau_\mathrm{KH}$ is much shorter---it is below $1$~min, more precisely $57$~s vs $370$~s in a single-twisted flux tube.  The wave phase velocity of the unstable kink mode is, however, with $66$~km\,s$^{-1}$ higher than that velocity in a single-twisted tube been equal to ${\cong}205$~km\,s$^{-1}$.

Our study shows that the KHI occurrence of the kink ($m =1$) mode primarily depends on the density contrast between the two media: the jet and its environment.  Here comes the big question: ``Which is the adequate approach of defining the density contrast: the standard one or that suggested by \cite{Paraschiv15}?''  The answer can only be obtained when we have more information about the parameters of an observationally recorded KHI, for example like that in a coronal mass ejection registered by \cite{Foullon11} and \cite{Ofman11}, namely observationally deduced wavelength $\lambda_\mathrm{KH}$, instability growth rate $\gamma_\mathrm{KH}$ and wave phase velocity $v_\mathrm{ph}$ to be compared with their values obtained from numerically derived growth rate and dispersion curves.  The Kelvin--Helmholtz instability is an important instability as we have mentioned in the Introduction section because it is capable of converting well-ordered flows, such solar EUV and X-ray jets, into more disordered, even turbulent, flows, which can lead to a heating of the solar atmosphere.

\begin{center}
Acknowledgments
\end{center}
The work of M.B.\ and I.Zh.\ was supported by the Bulgarian Science Fund under Indo-Bulgarian bilateral project DNTS/INDIA 01/7, and that of R.J.\ and R.C.\ by the Department of Science \& Technology, Government of India Fund under the project /Int/Bulgaria/P-2/12.  The authors are thankful to the \emph{Solar Dynamic Observatory}, the data from which are used in the present investigation.  We are also indebted to Dr.~Snezhana Yordanova for plotting one figure.

\bibliography{references}

\end{document}